\documentclass[preprint,aps,prl,epsf]{revtex4}   % makes power law references
\include{amsmath}
\include{epsfx}
\usepackage{graphicx}

\begin{document}
\title{{\bf Nanoindentation of virus capsids in a molecular model}}
\author{{\bf Marek Cieplak$^1$ and Mark O. Robbins$^2$}}

\address{
$^1$Institute of Physics, Polish Academy of Sciences,
Aleja Lotnik{\'o}w 32/46, 02-668 Warsaw, Poland \\
$^2$Department of Physics and Astronomy, Johns Hopkins University,
3400 North Charles Street\\
Baltimore, MD 21218, USA}

%$^*$Correspondence to: \\
%Marek Cieplak,\\
%Institute of Physics, \\
%Polish Academy of Sciences, \\
%Al. Lotnik\'ow 32-46  \\
%02-668 Warsaw, Poland\\
%Tel:  48-22-843-7001\\
%Fax:  48-22-843-0926\\
%E-mail: mc@ifpan.edu.pl

\vskip 40pt
%\noindent {\bf
%Keywords: 

\begin{abstract}
{A molecular-level model is used to study the mechanical response
of empty cowpea chlorotic mottle virus (CCMV) and
cowpea mosaic virus (CPMV) capsids.
The model is based on the native structure of the proteins that
consitute the capsids and is
described in terms of the C$^{\alpha}$ atoms.
Nanoindentation by a large tip is modeled as compression between
parallel plates.
Plots of the compressive force versus plate separation for CCMV
are qualitatively consistent with continuum models and
experiments, showing an elastic region followed
by an irreversible drop in force.
The mechanical response of CPMV has not been studied, but
the molecular model predicts an order of magnitude higher stiffness
and a much shorter elastic region than for CCMV.
These large changes result from small structural changes that
increase the number of bonds by only 30\% and would be difficult
to capture in continuum models.
Direct comparison of local deformations in continuum and molecular models
of CCMV shows that the molecular model undergoes a gradual symmetry breaking
rotation and accommodates more strain near the walls than
the continuum model.
The irreversible drop in force at small separations is associated
with rupturing nearly all of the bonds between capsid proteins
in the molecular model while a
buckling transition is observed in continuum models.
}
\end{abstract}

\maketitle

\section{Introduction}

Understanding the properties of self-assembled molecular structures
such as virus capsids is of great current interest.
Capsids are conglomerates of proteins that protect virus's genomes.
They are generally strong mechanically, especially when they envelope tightly
packed DNA at high internal pressures. For instance, in bacteriophage $\phi 29$, 
the pressure is of order 50 atm \cite{Evilevitch,Lenz,Lenz1}. Capsids that protect 
single stranded RNA have an order of magnitude lower pressures, 
as evidenced by Michel et al. \cite{Michel} for the 
cowpea chlorotic mottle virus (CCMV), but they are still quite sturdy.

One way of assessing the elastic
properties of capsids is by making nanoindentation measurements
with an atomic force microscope. In particular, it has been found \cite{Michel}
that the CCMV capsids with their RNA removed resist nanoindentation less
(by about 30\%) than their full counterparts. Both types of CCMV capsids
are highly elastic  -- the initial Hookean reversible regime persists up to
nanoindentions  varying between 20\% and 30\% of the diameter and applied
forces of 600 pN and 1000 pN for the empty and full capsids respectively
\cite{Michel,Wuite}.
These forces are noticeably larger than characteristic forces needed for a
mechanical unraveling of proteins. As an example, titin, which is a reasonably
strong protein, requires a force of about 204 pN \cite{R4,C6} to unravel
whereas the particularly strong scaffoldin protein c7A requires a force
of about 480 pN \cite{Valbuena}.

The sturdiness of the capsids comes from both the mechanical properties 
of the individual proteins and the way the proteins are bound together.
Capsids have a variety of shapes and sizes, 
but most have highly symmetric icosahedral structures.
Early models treated the capsid as a homogeneous spherical shell,
first in the thin shell limit \cite{Ivanovska04,Michel} and then including the
finite width of the shell and nonlinear response \cite{Gibbons07}.
The latter work showed that a Hookean model for finite width shells did not
reproduce the linear force displacement curve observed in experiments, while
standard nonlinear models for rubberlike materials did give a linear response.
The icosahedral symmetry of viruses has been included in the thin
shell limit \cite{Wuite,Gompper,Lenz,Lenz1} and using the mass density to determine
a spatially varying shell thickness \cite{Gibbons}, but with homogeneous
elastic response.

Many of the above models are roughly consistent with experimental behavior.
The force rises linearly with displacement for an extended range and then
drops suddenly, signaling an instability.
The linear region can be fit by assuming elastic moduli with a reasonable
magnitude, but the fact that different models can achieve this suggests
that the result is not sensitive to variations in local moduli and density
that may be present in capsids.
For example, one may expect that the number and geometry of bonds 
within proteins are different than those between proteins,
and this heterogeneity is difficult to capture in continuum theories.
Another issue is that the instability at higher forces in continuum
models is associated with a buckling instability
without any breaking of molecular bonds.
The latter may be important in real capsids, but is difficult to parameterize
in continuum models.
More detailed models are needed to assess potential variations in elastic
properties within the shell and the potential role of bond breaking.

All-atom simulations are an appealing alternative, but are limited to relatively
short indentation times of order 1ns compared to experimental times of 10ms or more.
Studies of mechanical properties have included an all-atom
analysis of the low frequency vibrations in the satellite tobacco necrosis 
virus \cite{STNV} and an elastic network model analysis of eigenmodes in
CCMV \cite{Brooks1} and HK97 viruses \cite{Chirikjian, Brooks2}.
Recent simulations by Zink and Grubmuller \cite{zink09} have examined
indentation of the southern bean mosaic capsid using
an atomistic treatment with surrounding water.
The force on a small ($\sim 1$nm) sphere representing the tip was measured as a
function of the distance from the center of mass.
They find a rather short elastic region (~1nm) followed by
a rapid drop in force.
The instability is associated with bond breaking, but the bonds rapidly
reform as the sphere passes through the capsid wall and into the interior.
This indicates the potential importance of factors that are not included
in continuum models.
Both the yield elastic stiffness and yield stress are about an order
of magnitude higher than experiment, but vary logarithmically with rate.
The yield stress often varies logarithmically with rate in thermally
activated systems \cite{eyring}, but it is unusual to find this variation in
the stiffness.
This suggests that the simulation rates are higher than the
structural relaxation rate of the virus and that new behavior may occur
at lower velocities.

In this paper we consider a model that is
less computationally intensive, but includes coarse-grained structural
information.
The model is used to explore the effect of molecular structure on
nanoindentation of CCMV and cowpea mosaic virus (CPMV),
whose symmetry is that of a rhombic tricontahedron \cite{CPMV1,CPMV2}.
Capsids in both systems contain of order 300 000 heavy atoms, which makes it 
difficult to study large mechanically-induced conformational changes with
all-atom simulations.
Instead we use structure-based coarse-grained 
molecular dynamics models that
take the experimentally determined native structure as an input to
derive the effective Hamiltonian in a phenomenological way. Our 
implementation builds on the detailed description presented in 
refs. \cite{Hoang2,Hoang,JPCM} and this approach has been successful in
describing both folding and mechanically-induced unfolding of individual
proteins \cite{Hoang2,JPCM}.
The coarse-graining reduces the number of effective protein atoms by an order of
magnitude.
Treating solvent molecules implicitly through
random force and damping terms leads to even greater economies
by eliminating atoms and increasing the effective time step.

We find that
the mechanical response of the molecular model of CCMV is qualitatively
consistent with experiments and continuum calculations,
with a long elastic region followed by a sudden drop in force.
CCMV and CPMV have similar radii, widths and densities and thus would 
be expected to have similar responses in continuum models.
In contrast, the coarse-grained model predicts that CPMV is an order
of magnitude stiffer and breaks at a higher force and smaller
displacement.
These large changes in mechanical response result from small
changes in bonding, suggesting that the mechanical
response of virus capsids may be highly variable.
Experiments on CPMV would provide a test of this prediction.

The local deformations in the molecular and continuum models \cite{Gibbons}
for CCMV are compared directly.
One significant difference is that the molecular model undergoes
a gradual and reversible rotation in the elastic region.
More strain is also accommodated near the walls in the molecular model.
Dramatic differences between the models are observed at the end of the
elastic region where the response becomes irreversible.
In the continuum model, nonlinearities are associated with buckling.
In the molecular model, nearly all of the interprotein bonds
break at the end of the elastic region.
This breaking transition appears to be thermally activated
and its location depends on trajectory.
The mean force and displacement at rupture decrease with
decreasing deformation rate, because there is more time for
thermal activation.

\section{The model systems}

CCMV is a member of the bromovirus
group of the {\it Bromoviridae} family. It consists of a single-stranded
RNA molecule encapsulated by a 180-protein icosahedral capsid.
The native CCMV capsid is stable around pH 5.0, where it is 
characterized by the outer and inner diameters of 286 and 195 {\AA}
respectively \cite{structure_CCMV}. At pH 7.0 and low ionic strength,
the capsid adopts its swollen form in which its linear sizes are
increased by about 10 \%.
Nanoindentation studies at pH 5 yield a
spring constant of about 0.15 N/m \cite{Wuite,Michel}.
At pH 6, which is half-way to the
situation in which the swollen form arises, the spring constant falls
by about a factor of about 3 compared to the native form \cite{Wuite}.
Furthermore, there is no longer a sudden drop
in force that would indicate mechanical instability.

The symmetries of the native capsid are those
of the fullerene C$_{60}$.
The proteins form a closed shell that has the symmetry of 
a truncated icosahedron comprised of 32 faces (making 90
edges and 60 vertices). 12 of these
faces are pentagons and 20 are hexagons. A pentagon consists of
five icosahedral asymetric units ('kites') and a hexagon contains
six such units. Molecularly, a pentagon corresponds to a
pentamer and a hexagon to a hexamer.
%MR Not sure what this sentence above means "Molecularly"? but OK.
There are thus altogether 180 units, each containing a single
molecule referred to as a chain. Even though the chains are
sequentially identical, one distinguishes three kinds of chains: 
A, B, and C. The pentamers contain only the A-type chains whereas the
hexamers have three chains of type B and three chains of type C.
All chains contain
190 amino acids in a $\beta$-barrel fold, but their conformations
have not been determined fully through X-ray crystallography.
In particular, the
locations of the first 41 amino acids in chain A and the first
26 in B and C are not known.
These undetermined segments of the proteins are presumably in more
random or mobile regions that do not contribute to the scattering
or structural integrity.
In our model, we make use only of the known parts of the structure
and the whole capsid is represented by 28620 C$^{\alpha}$ atoms
\cite{structure_CCMV,vdb}.

The conformation of the swollen form of CCMV at pH 7 has
been determined through cryo-electron microscopy \cite{structure_CCMV},
but at a much poorer resolution than the structure at pH 5.
The resolution is comparable to the separation between C$^{\alpha}$ atoms
that defines the structure in our coarse-grained model.
We constructed a model based on this structure, but
found that it was not mechanically stable even at $T=0$.
This could imply that atoms that are misplaced or 
missing from the structure are crucial to stability.
Note that the virus disassembles at slightly higher pH and may be nearly
unstable at pH 7.
We are not aware of indentation experiments at pH 7.
Given the large reduction in measured stiffness between pH 5 and pH 6
\cite{Wuite}
and the fact that swelling of the structure mainly occurs above pH 6,
it would be interesting to extend the experiments to the fully swollen regime.
Structural information at pH 6 would also allow us to compare our
model to experiments in the partially swollen state.

CPMV belongs to the picornavirus superfamily and
can be represented by a rhombic triacontahedron with 30 rhombic faces that 
make 60 edges and 32 vertices \cite{CPMV1,CPMV2}.
The capsid is assembled from two viral proteins -- subunits S and L --
which form three $\beta$-sandwich domains, A, B, and C in an asymetric unit.
They contain 189, 187, and 182 amino acids respectively.
Domains B and C form hexamers and each L subunit comprises B- and
C-type domains.
The A-domains form pentamers. Altogether, CPMV contains 60 units corresponding
to the total of 33 480 C$^{\alpha}$ atoms.
In contrast to CCMV, the positions of all amino acids are determined.

The angle-averaged radial distributions of the C$^{\alpha}$ atoms in the two
capsids in their native states are shown in Figure \ref{radial}.
The average radial position is 119.6 {\AA} in CCMV and  124.3 {\AA} in CPMV.
The distributions of radii are also comparable,
with total widths of about 55 {\AA}
and rms variations of 11.13 {\AA} and 10.54 {\AA}, respectively.
In continuum theory the stiffness associated with compressing a thin shell
scales as $h^2/R$ where $h$ is the thickness and $R$ the radius \cite{Michel}.
Given the similarities in the number and radial distribution of C$^{\alpha}$
atoms one would expect similar stiffnesses, but we will
see that this is not the case.  

The coarse-grained models of the capsids generalize our previous approach
for large conformational changes in single 
proteins \cite{Hoang2,Hoang,JPCM}.
We take the native coordinates of all heavy atoms in all capsid proteins
from the structure files and assign 
van der Waals spheres to the atoms (enlarged by a factor 
that takes into account attraction \cite{Tsai}).
If the spheres belonging to amino acid $i$, as counted along the "global"
sequence, overlap the spheres related to amino acid $j$ then we declare
existence of a native contact between amino acids $i$ and $j$.
Most of the native contacts are between acids in the same protein, but there
are also contacts that link separate proteins.
Contacts corresponding to the i,i+2 interaction within a protein are
discarded \cite{models} as they usually
correspond to dispersive interactions which are much weaker than
the hydrogen bonds. Altogether, there are 62 426 native contacts in the
model CCMV and 90 420 contacts in the model CPMV. 
This 45\% increase is much greater than the change in the number of C$^\alpha$
atoms or their radial distribution,
and may be an important factor in explaining why we find
a much greater stiffness for CPMV than CCMV.

The interactions between amino acids that form native contacts are described
by the Lennard-Jones (LJ) potential
$V(r)=4\;\epsilon [\left(\frac{\sigma_{ij}}{r_{ij}}\right)^{12}-
\left(\frac{\sigma_{ij}}{r_{ij}}\right)^{6}]$, where
$r_{ij}$ is the distance between the C$^{\alpha}$'s in amino acids $i$ and $j$,
and $\sigma _{ij}$ is determined for each pair to ensure that the potential
minimum coincides with the experimentally measured native distance.
The binding energy parameter $\epsilon$ can be determined by comparing
theoretical
and experimental data for stretching of single proteins \cite{models}.
Our latest estimate \cite{plos} is that $\epsilon$ is of order 110 pN{\AA}
(see also ref. \cite{Marszalek}).
Interactions along the backbone of each protein are described with a
harmonic potential with spring constant $50 \epsilon/$\AA$^2$.
This is almost an order of magnitude stiffer than the LJ interaction
and ensures that the backbone bonds do not break.

The interactions between atoms that do not form native contacts
are represented by a purely repulsive potential obtained by truncating
the Lennard-Jones potential at its minimum at 4 {\AA}.
The potential number of non-native contacts is
of order 5$\times 10^8$.
To avoid calculating all these separations at each step,
we make a Verlet list \cite{Allen}.
The list comprises all C$^{\alpha}$ 
atoms that are less than 12 {\AA} apart from another C$^{\alpha}$ atom
and is made every 2000 time steps.

Langevin noise and damping terms
are introduced to provide an implicit solvent \cite{Grest}.
The amplitude of the noise is determined by the temperature, $T$. All
simulations described below were done at $k_BT=0.3 \epsilon$,
where $k_B$ is the Boltzmann constant.
Estimates of $\epsilon$ for our model place $k_BT/\epsilon$ between
0.3 and 0.4 at room temperature.
We find similar behavior for
indentation over this range of temperatures
although there are small changes in quantitative values.
The model is simulated in the overdamped limit and
the characteristic time scale $\tau$ is set by the damping rate
$2m/\tau$ in the Langevin thermostat, where $m$ is the average mass of
an amino acid.
%MR Need to define what mass is, right?
%each C$^\alpha$.
For proteins in water $\tau$
is of order 1 ns as argued in refs. \cite{Veitshans,Peclet}.
The time step used to integrate the equations of motion
with a fifth order predictor-corrector algorithm \cite{Allen} is
$\tau/200$.

Nanoindentation is implemented by placing the capsid between two repulsive 
parallel plates.
This geometry corresponds to the limit of a tip with large radius and
has been considered in continuum calculations.
The experimental geometry is often a flat substrate with a tip whose
radius is of the same order as the virus, but continuum results show
that this has little effect on the stiffness.
The plate potential scales as $h^{-10}$ \cite{Steele},
where $h$ is the separation between the plate and an amino acid.
In the initial state, the plates are far enough apart that they
do not interact with the capsid.
The plates are then brought together by increasing
the speed of both plates symmetrically to a value $v_p$
over 2000 $\tau$.
In most cases $v_p=0.0025$ {\AA}$/\tau$ giving
a combined speed of 0.005 {\AA}/$\tau$
that was found to be slow enough to produce quasistatic
results in studies of protein stretching
\cite{JPCM,models}. 
We will also present results at twice this velocity and half this velocity
and refer to these as fast and slow compression, respectively.
While these velocities ($\sim 500$ $\mu$/s)
are higher than experimental velocities (0.1 to 10 $\mu$m/s),
they are slow enough that stress can be transmitted across the capsid
before the separation has changed substantially.

We determine total force from the capsid on each plate and then
average the magnitude of these forces
to obtain the total compressive force that would
be measured by an AFM.
The inset in Figure \ref{basic} shows that deviations between
the forces on the two plates are similar to temporal fluctuations
in each.
Larger deviations with longer duration are observed if the velocity
is too high for stress equilibration.
In this and subsequent figures,
the mean force is plotted
as a function of the plate-to-plate separation, $s$.
While the plate potential has a finite range, it changes so rapidly
that the effective space available to the capsid is only slightly smaller
than $s$.

Since capsids are not spherical, their elastic properties should depend on their
orientation relative to the compressive axis.
We consider two orientations here.
Most results are for the case where one of the symmetrically
equivalent 2-fold axes is perpendicular to the plates.
This corresponds to the $z$-axis as defined in the structure file \cite{vdb}.
The second situation has a 3-fold axis normal to the plates.
This corresponds to the [1,1,1] direction in the structure file and
is denoted by [1,1,1].
Note that these orientations represent the initial configuration and
the capsids can rotate under the applied load.
The observed rotations are relatively small and are discussed further below.

\section{Results}
\subsection{Force-Separation Curves}

\begin{figure}
\epsfxsize=6in
%\centerline{\epsffile{upbczar.eps}}
\centerline{\epsffile{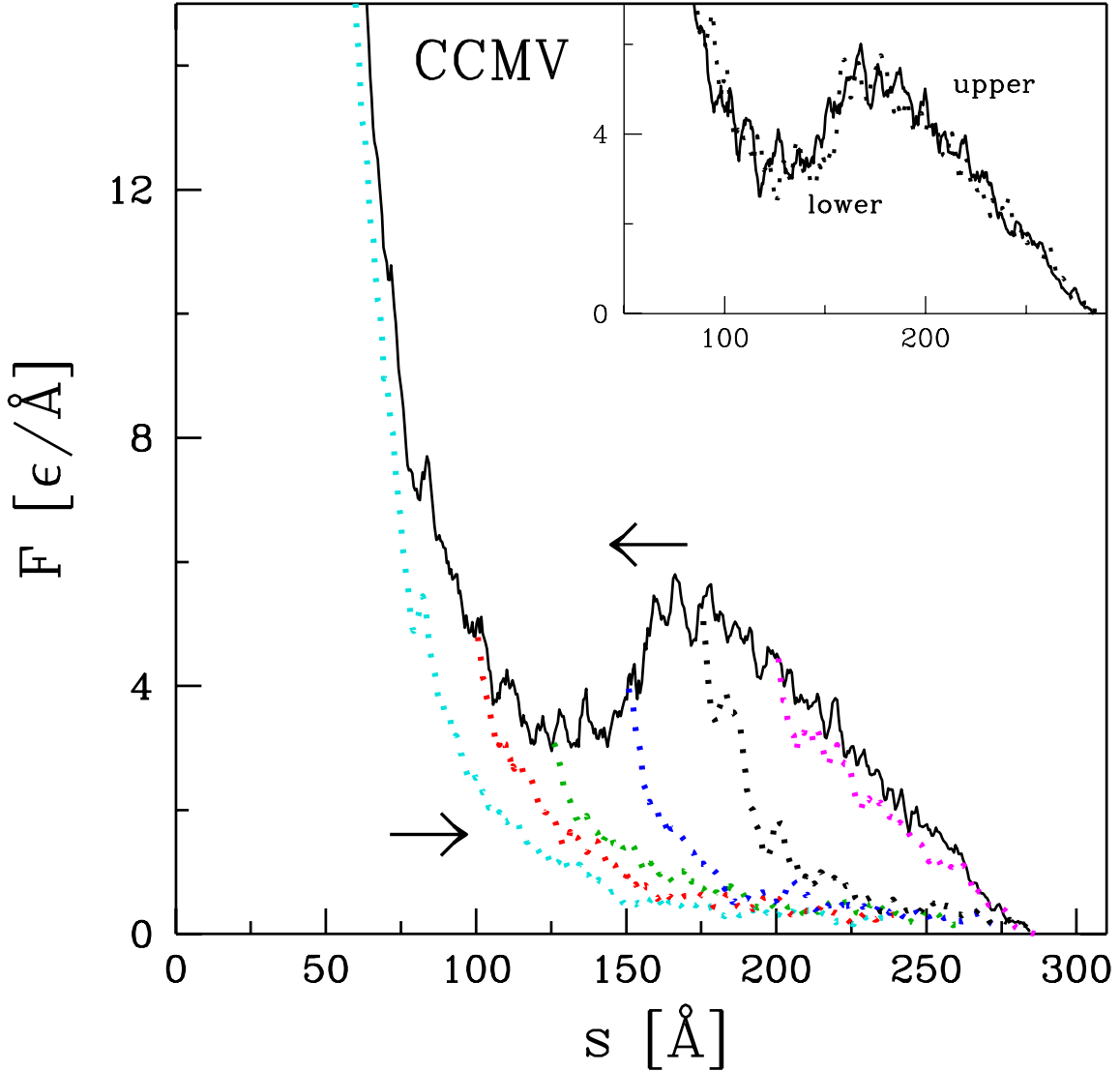}}
\vspace*{3cm}
\caption{Force on confining plates from CCMV capsid atoms as a function of the
plate-plate separation in one trajectory corresponding to $v_p$=0.0025 {\AA}/$\tau$.
The solid line corresponds to nanoindentation. The dotted lines show the force
observed when the plates move back at the same speed after reaching the separation 
indicated by detachment of dotted lines from the solid line.
The inset shows the separate forces acting on the top (solid)
and bottom (dashed) plates. 
The average of these forces is the solid line in the main figure.
}
\label{basic}
\end{figure}

Figure \ref{basic} shows an example of a nanoindentation trajectory
for the CCMV capsid.
No force is observed for separations greater than 284 {\AA}.
The force then rises linearly as $s$ decreases to about 175 {\AA}.
This corresponds to a compression by about 35\% from the point of
first contact, a surprisingly large range of Hookean response,
but comparable to what is observed in experiments on capsids.
Note that experimental data are typically plotted vs. the displacement
of the tip towards the substrate, which becomes more positive as $s$ decreases.
One must also be careful to remove the compliance of the AFM cantilever
in comparing to experimental numbers.

\begin{figure}
\epsfxsize=6in
%\centerline{\epsffile{cpgrub.eps}}
\centerline{\epsffile{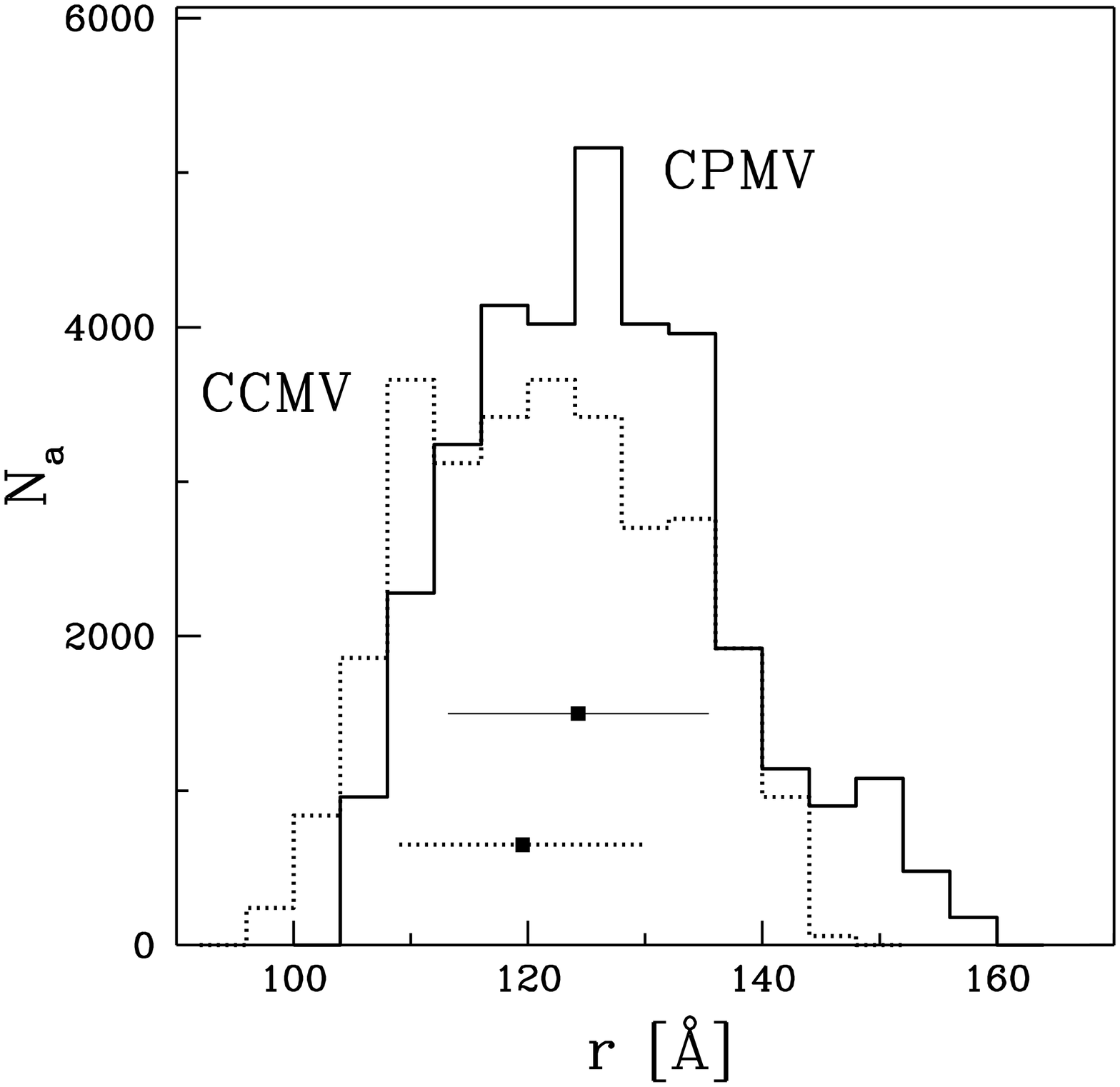}}
\caption{The radial distributions of the C$^{\alpha}$ atoms in the
CCMV (dotted line)
and CPMV (solid line)
capsids. The solid squares indicate the mean radii
and the horizontal lines show the sizes of the
standard deviations.}
\label{radial}
\end{figure}

At the end of the Hookean regime the force
drops rapidly by about 40\%.
A comparable drop is observed in experiments \cite{Michel,Wuite}.
At still smaller separations the force rises rapidly.
Given the observed width of the radial distributions in
Figure \ref{radial}, it is not surprising to find that the
top and bottom of the
virus are in direct contact at these small separations, leading
to strong repulsive forces.
 
In the regime where the force rises linearly,
the nanoindentation 
process is nearly reversible: An inversion of the direction of motion
of the plates results in a force which returns along the same
straight line.
Once the capsid has been compressed to or beyond the sudden drop
in force, the deformation is irreversible.
To illustrate this behavior we reversed the velocity of the walls
after compressing to 50, 100, 125, 150, 175 and 200 {\AA}
and the results are plotted as dotted lines in Fig. \ref{basic}.
The behavior is only reversible for $s$=200 {\AA}.
The small drop at $s$ just greater than 175 {\AA} is already
signalling the onset of plastic deformation that leads to
extremely irreversible behavior.
The nature of this plastic deformation is discussed further below.

\begin{figure}
\epsfxsize=6in
%\centerline{\epsffile{jbczar.eps}}
\centerline{\epsffile{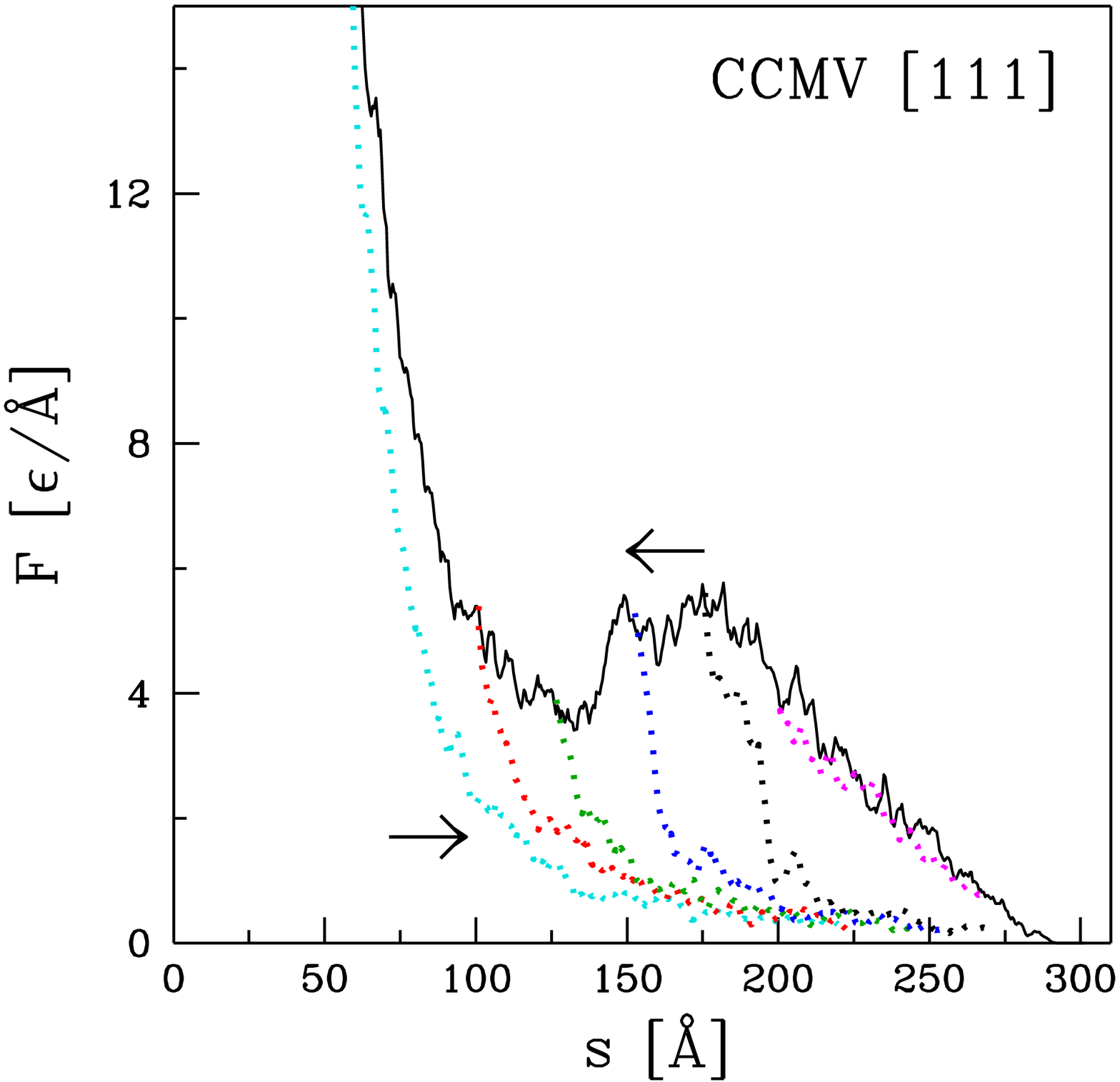}}
\vspace*{3cm}
\caption{Similar to Figure \ref{basic} but for nanoindentation along the
[1,1,1] direction.}
\label{basic111}
\end{figure}

Figure \ref{basic111} shows that nanoindentation of a capsid with the
[1,1,1] orientation produces very similar results.
The diameter along this orientation is slightly bigger, so the force starts
at a larger $s$, but values of $F_m$ and $k$ are very similar to those
in Figure \ref{basic}.
The onset of irreversibility also coincides with the end of the Hookean
regime.
Note that while the capsid was free to rotate,
the capsid remained in roughly its initial orientation
in both Figures \ref{basic} and \ref{basic111}.

\begin{figure}
\epsfxsize=6in
%\centerline{\epsffile{sbczar.eps}}
\centerline{\epsffile{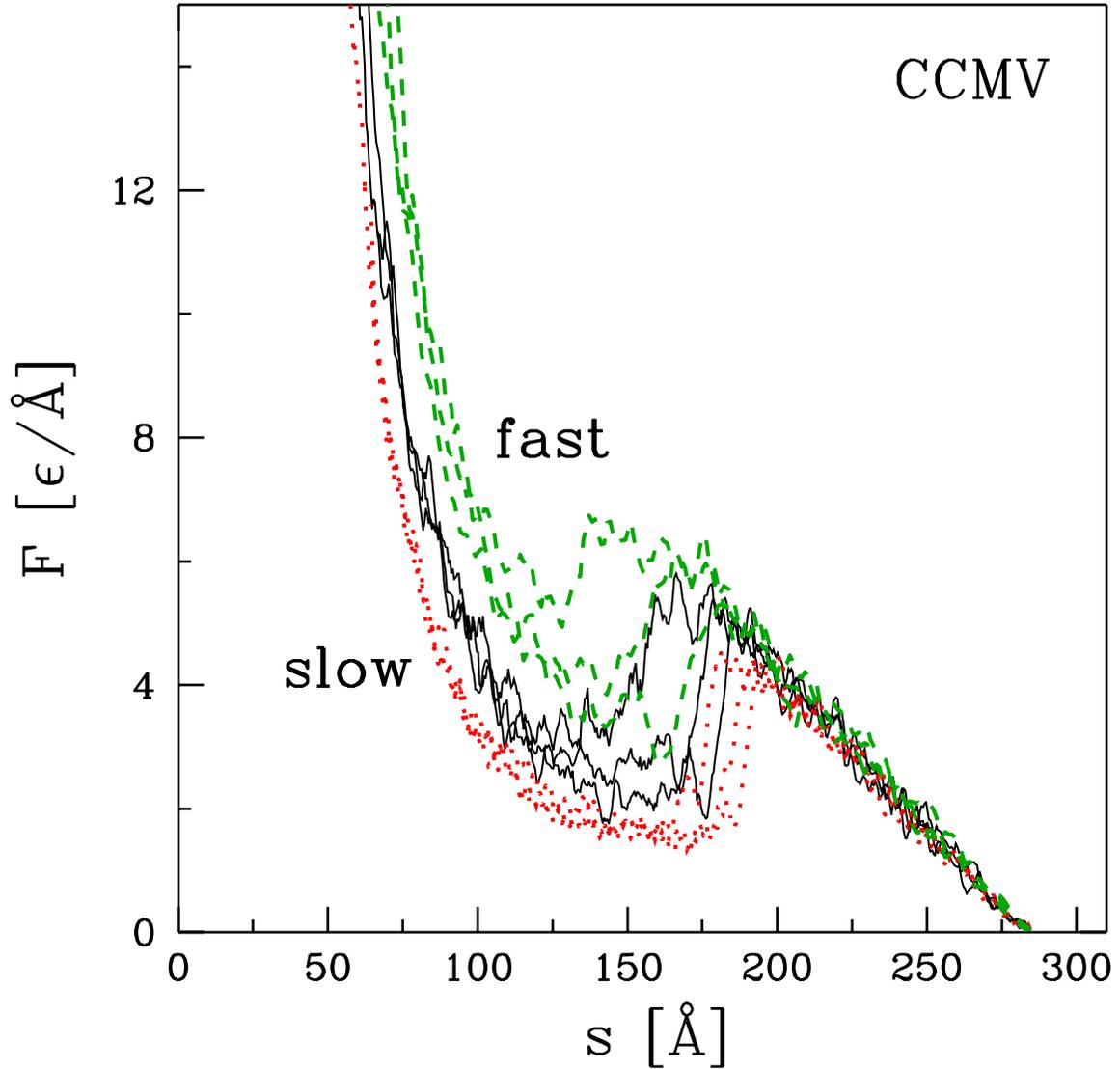}}
\vspace*{3cm}
\caption{Variation of the force on a CCMV capsid with separation for
different trajectories and compression speeds.
The solid lines in the center
correspond to three trajectories at $v_p=0.0025$ {\AA}/$\tau$.
The three dotted and dashed lines correspond to trajectories at
half and twice this speed, respectively.
}
\label{speed}
\end{figure}

The velocity dependence of our results is illustrated in Figure \ref{speed}.
Three different realizations of the system are shown for the velocity
considered above, as well as for velocities that are slower and faster
by a factor of two.
The different realizations at a given speed
exhibit essentially the same behavior
at large and small separations, but the onset of irreversibility
occurs at different forces and separations.
This variability suggests that the force drop represents an activated
transition from the stable linear regime to another state.
The rate dependence is also consistent with this conclusion.
In particular, thermal fluctuations have more time to activate the transition
as the velocity decreases, and thus the average
force at the transition decreases with decreasing velocity.
The nature of the transition is discussed further below.

The stiffness of the capsid in the linear regime is relatively insensitive
to velocity, with fluctuations comparable to variations with speed.
The results are consistent with a very small (5\%) increase in stiffness
with speed.
A weak linear dependence is expected since atoms are able to relax
more closely to the free energy minimum as the velocity decreases.
A stronger velocity dependence is observed at very small $s$.
This is the regime where the top and bottom of the capsid are in hard
sphere contact.
The rapid rise of the repulsive force with decreasing separation makes
the system more sensitive to unrelaxed positions in this regime.

\begin{figure}
\epsfxsize=6in
%\centerline{\epsffile{bszar.eps}}
\centerline{\epsffile{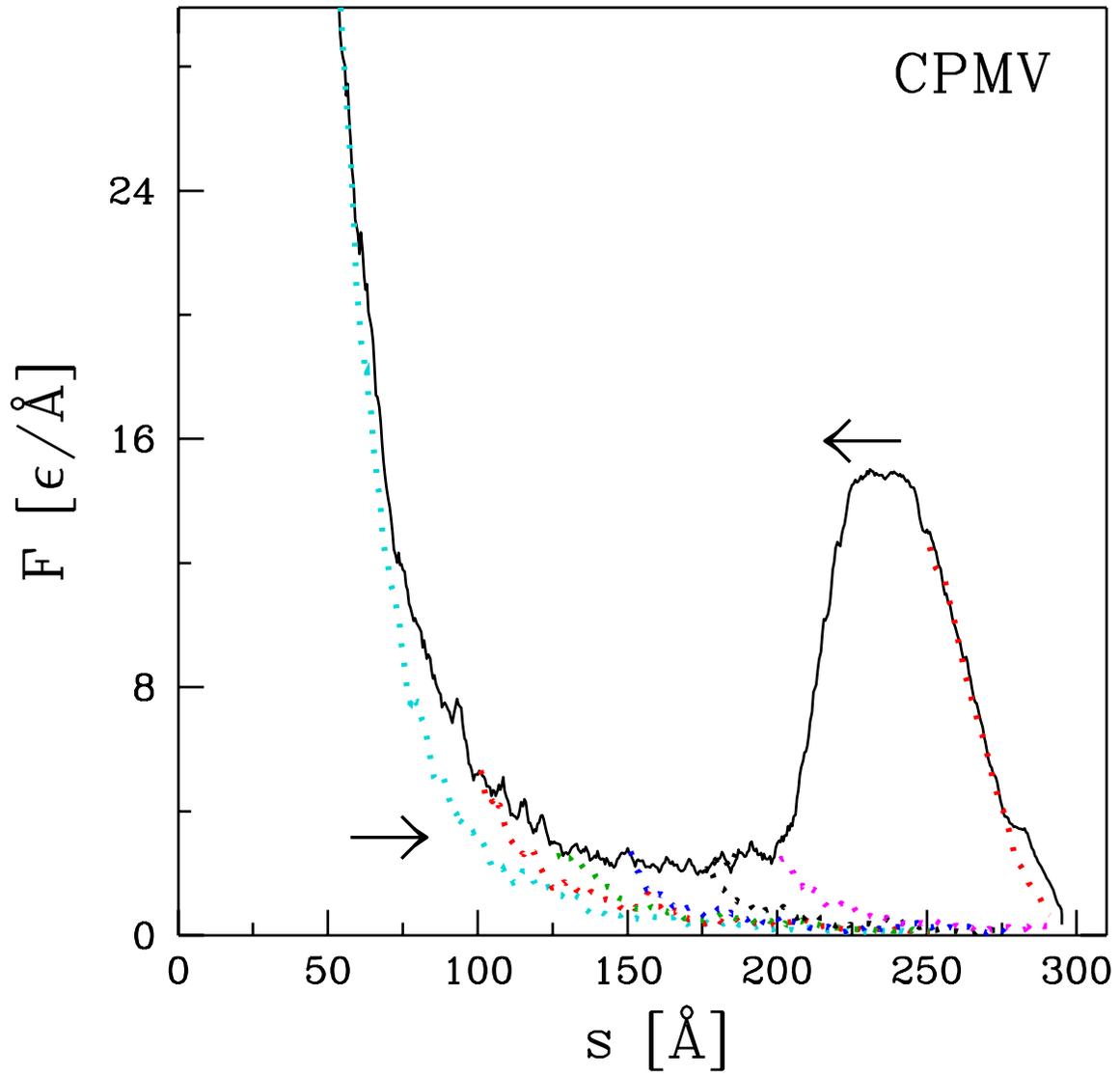}}
\vspace*{3cm}
\caption{Similar to Figure \ref{basic} but for CPMV. }
\label{bascpmv}
\end{figure}

\begin{figure}
\epsfxsize=6in
%\centerline{\epsffile{jbszar.eps}}
\centerline{\epsffile{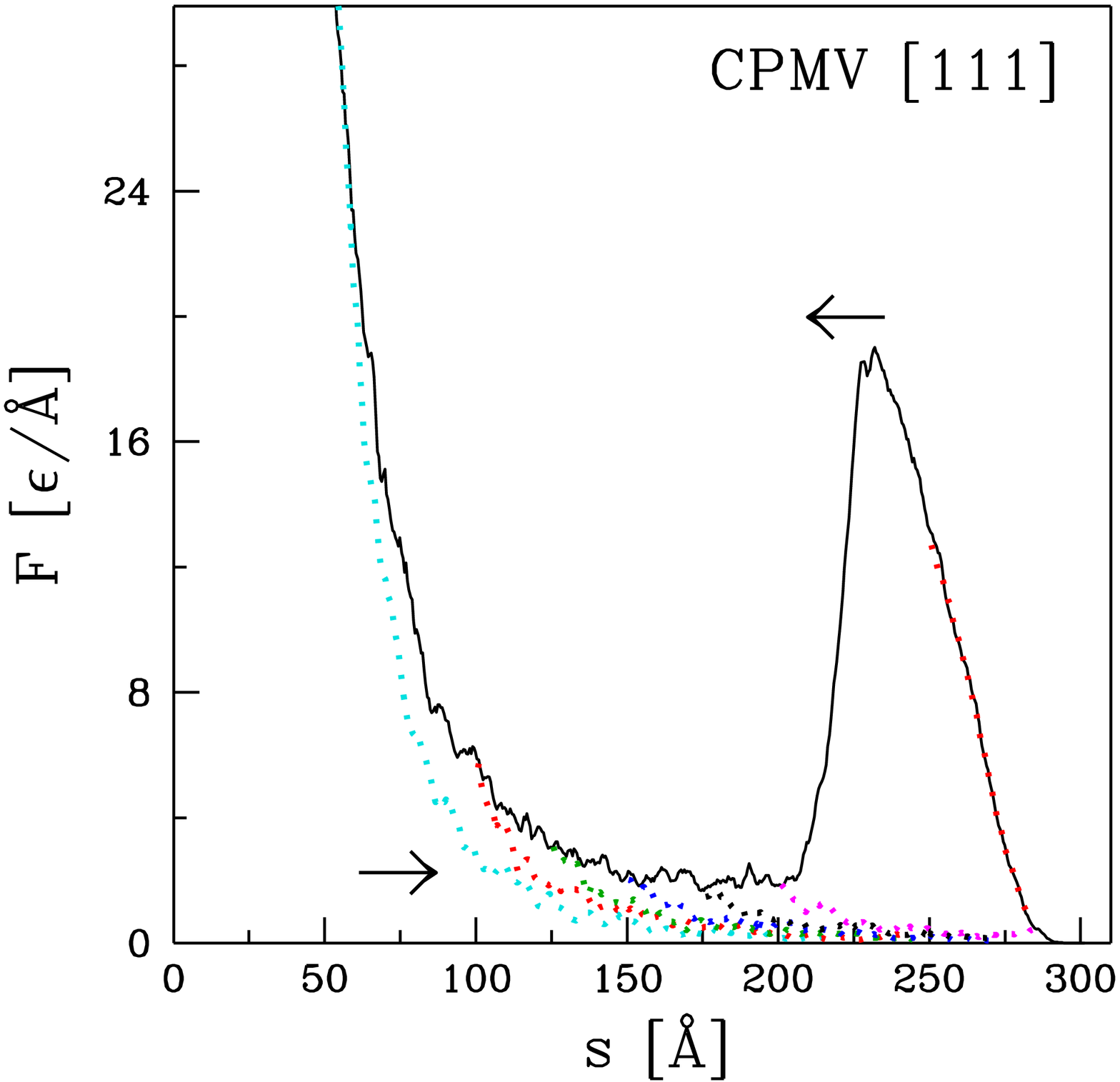}}
\vspace*{3cm}
\caption{Similar to Figure \ref{bascpmv} but for nanoindentation
along the [1,1,1] direction.} 
\label{bascpmv111}
\end{figure}

The CPMV capsid is found to behave in a qualitatively similar way. 
As shown in Figure \ref{bascpmv}, there is a regime where the force
rises linearly, followed by a sudden drop in force.
The force-separation curve is reversible in the early stages of
compression, but becomes irreversible after the end of the linear
regime.
The main differences between the capsids are that CPMV is much stiffer
in the linear regime and fails at a higher yield stress and smaller
degree of compression.
The peak force $F_m \sim 15$ $\epsilon$/{\AA} is a factor of three
larger than for CCMV, while the
effective spring constant, $k\sim 0.5$ $\epsilon$/{\AA}$^2$, is an order
of magnitude higher.
Changing the orientation of the capsid to [111] produces similar results
(Figure \ref{bascpmv111}).
The peak force is slightly larger, but the difference is within the
range of fluctuations between different runs.

The large difference in stiffness of the two viruses is surprising.
Continuum theory predicts that $k=2.25 E h^2/R$ 
where $E$ is the elastic modulus, $h$ the shell thickness and $R$
the radius \cite{Michel}.
As shown in Fig. \ref{radial},
CCMV and CPMV have similar radii $R\sim 120$ {\AA} 
and thicknesses $h \sim 20$ \AA.
They also have the same interaction energies within our model,
and similar numbers of amino acids.
Yet the measured stiffnesses would imply elastic moduli that
differ by an order of magnitude:
0.007 $\epsilon $\AA$^{-3}$ and 0.07 $\epsilon$\AA$^{-3}$
for CCMV and CPMV, respectively.
To put these values in perspective,
they are
one to two orders of magnitude
smaller than the elastic modulus of a Lennard-Jones
crystal with the same interactions
$E$=55 $\epsilon/\sigma^3$ = 1.2 $\epsilon/$\AA$^3$.  
This large reduction must reflect the fact that the number of bonds constraining
the motion of each segment of the protein is relatively small.

Atoms in an fcc crystal have $z=12$ nearest neighbors or 6 bonds per atom.
CCMV and CPMV have only 2 or 3 native contacts per
atom, respectively.
Including the neighbors along the backbone of the protein,
this corresponds to about 6 neighbors for CCMV and 8 for CPMV.
The minimum number of neighbors required for stability in 3 dimensions is
$z=4$, and studies of rigidity percolation indicate that the elastic
modulus is a strong function $z$.
It seems likely that this explains why increasing the number of
native contacts per atom by only 45\% (and total bonds per atom by 30\%)
can increase the effective
stiffness of CPMV by an order of magnitude.
The observation that small changes in protein binding can have a
dramatic effect on stiffness suggests that further experimental
studies may reveal large variations between capsids.
We are not aware of any nanoindentation studies of CPMV, but it would be
interesting to see if the trends observed in our model are reproduced
in experiment.

\subsection{Changes in bonding }

\begin{figure}
\epsfxsize=6in
%\centerline{\epsffile{brontact.eps}}
\centerline{\epsffile{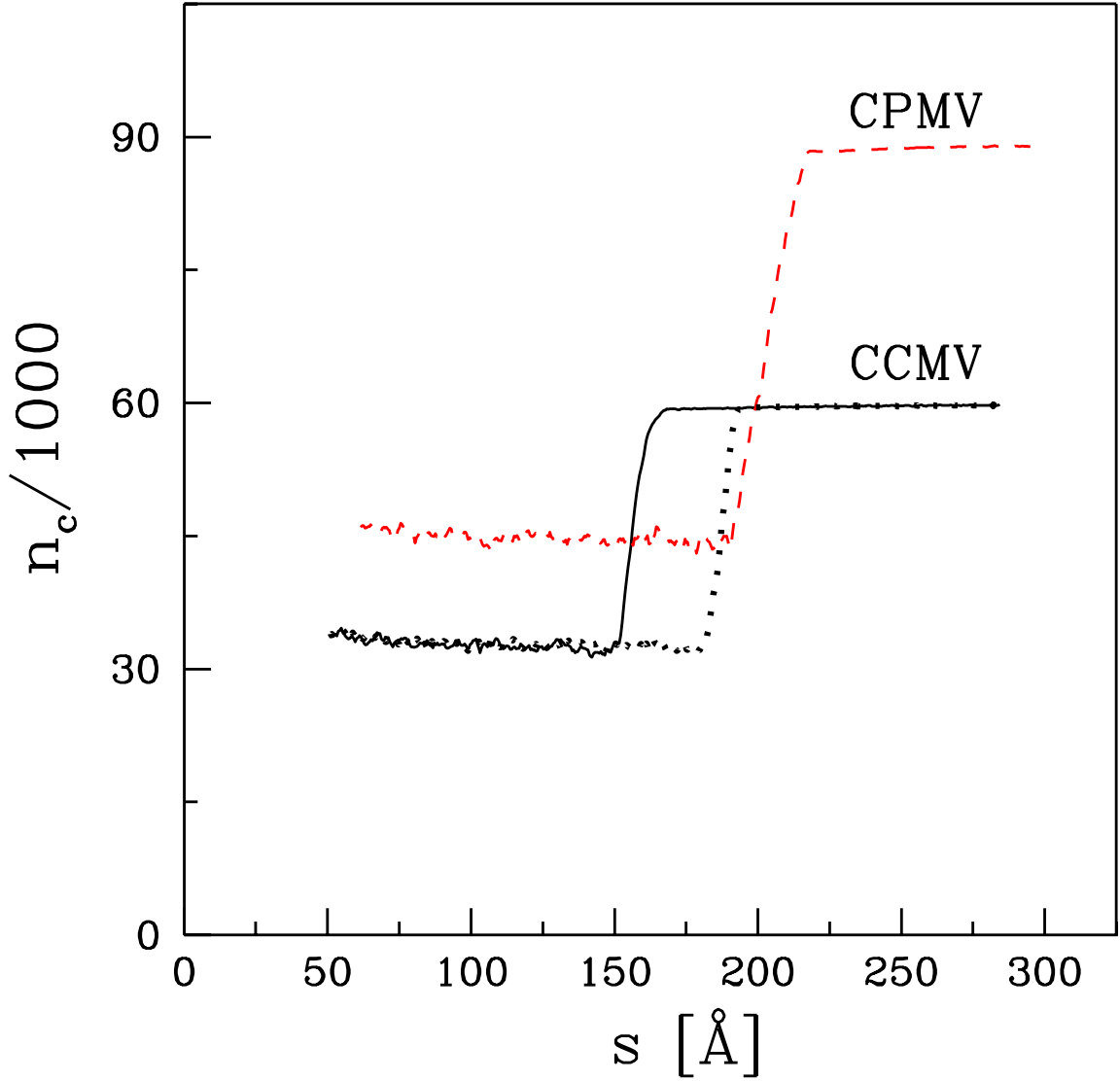}}
\vspace*{3cm}
\caption{The number of native contacts that remain unruptured as a function of
$s$ for CCMV (solid line) and CPMV (dashed line) in trajectories
corresponding to those shown
in Figures \ref{basic} and  \ref{bascpmv}, respectively.
A native contact is considered unruptured if the two amino acids
are within a distance of 1.5 $\sigma _{ij}$.
The dotted line is for compression of CCMV at half the speed and corresponds
to the lowest trajectory
in Figure \ref{speed}.}
\label{contacts}
\end{figure}

The onset of irreversibility in both CCMV and CPMV is associated with
rupture of native contacts rather than a buckling transition like that
observed in elastic shell models \cite{Gibbons,Michel, Gibbons07}.
Figure \ref{contacts}
shows the total number, $n_c$, of unbroken native contacts
($r_{ij}< 1.5 \sigma _{ij}$) as a function of $s$. 
In the initial stage of compression, the number of unbroken contacts remains
essentially constant.
Thermal fluctuations cause a very small number (~0.1\%) of bonds
to fluctuate in and out of contact, but the structure remains intact.
This behavior extends to the end of the Hookean regime where deformation
becomes irreversible.
At this point $n_c$ drops sharply
and then saturates as the virus is compressed further.
About half the contacts break during this transition in both
CCMV and CPMV.

As seen in the force plots described in the previous section,
the onset of irreversibility depends on thermal fluctuations
and tends to occur at larger $s$ for slower compression.
The drop in $n_c$ always coincides with the onset of irreversibility
and shifts in the same way.
The dotted line in Fig. \ref{contacts} illustrates an earlier onset
of bond rupture in CCMV at a lower rate.
Note that the values of $n_c$ for the different rates coincide before
and after rupture occurs.
Only the point at which thermal fluctuations lead to bond breaking
changes.

Figure \ref{contacts} shows the total number of contacts,
but it is interesting to examine intra and inter-protein bonds separately.
For the case of CCMV there are three times as many intraprotein contacts
(not counting bonds along the backbone of the protein) as interprotein
contacts: 47166 vs. 15300.
When bonds begin to break, the number of broken intraprotein bonds
is slightly higher, but the fraction of broken bonds is dramatically different.
By $s=149$ {\AA} over 95\% of the interprotein bonds are broken, while more than
half the intraprotein bonds remain.
This striking result indicates that rupture of the capsid occurs mainly along
the joints between proteins, something that would be difficult to capture
with homogeneous continuum models.

Interprotein bonds have the same energy parameters as intraprotein bonds,
but equilibrium simulations also provide evidence that the joints
where proteins meet are weaker.
The mean squared variation in the length of native contacts in
equilibrium should be given by the thermal energy divided by an
effective spring constant that is influenced by
bonds to surrounding atoms.
For CCMV,
the ratio of the mean squared fluctuation to the equilibrium bond length
is more than 80\% larger for interchain interactions
than for intrachain interactions, implying that the interprotein
regions are roughly half as stiff as intraprotein regions.
This greater fluctuation indicates that there are fewer reinforcing
contacts to limit deformation and prevent rupture along the borders
of proteins.

In CCMV, proteins fold into kite shaped pieces and 
three kites (A, B, and C chains) combine
to form triangles that tile the surface \cite{structure_CCMV}.
There are no intrachain bonds connecting the kites.
In CPMV, three trapezoids combine to form triangles.
Two trapezoids on adjacent triangles are formed from a single
molecule, providing direct intrachain links \cite{structure_CCMV}.
Given the mode of rupture,
it seems likely that these intrachain bonds
between triangles
play an important role in the high stiffness and rupture force of CPMV.

\subsection{Structural Changes in CCMV}

\begin{figure}
\epsfxsize=6in
%\centerline{\epsffile{birasm.eps}}
\centerline{\epsffile{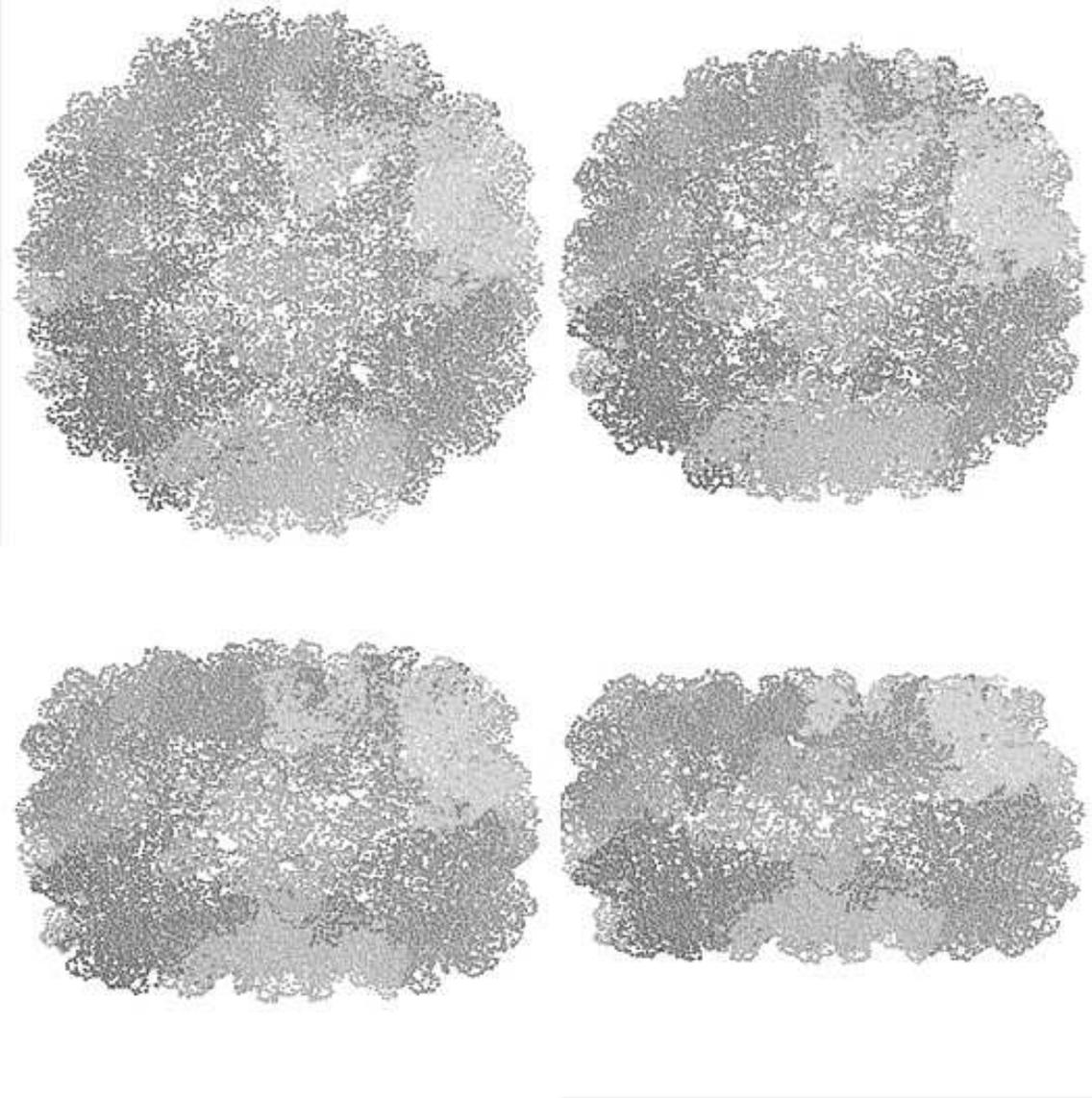}}
\vspace*{3cm}
\caption{The structure of the CCMV capside with all C$^{\alpha}$ locations shown
at $s$=284 {\AA} (top left), 239 {\AA} (top right), 189 {\AA} (bottom left, 
and 164 {\AA} (bottom right).  The smallest value of $s$ corresponds to the
start of the nonlinear regime.}
\label{pictures}
\end{figure}

\begin{figure}
\epsfxsize=6in
%\centerline{\epsffile{bitop.eps}}
\centerline{\epsffile{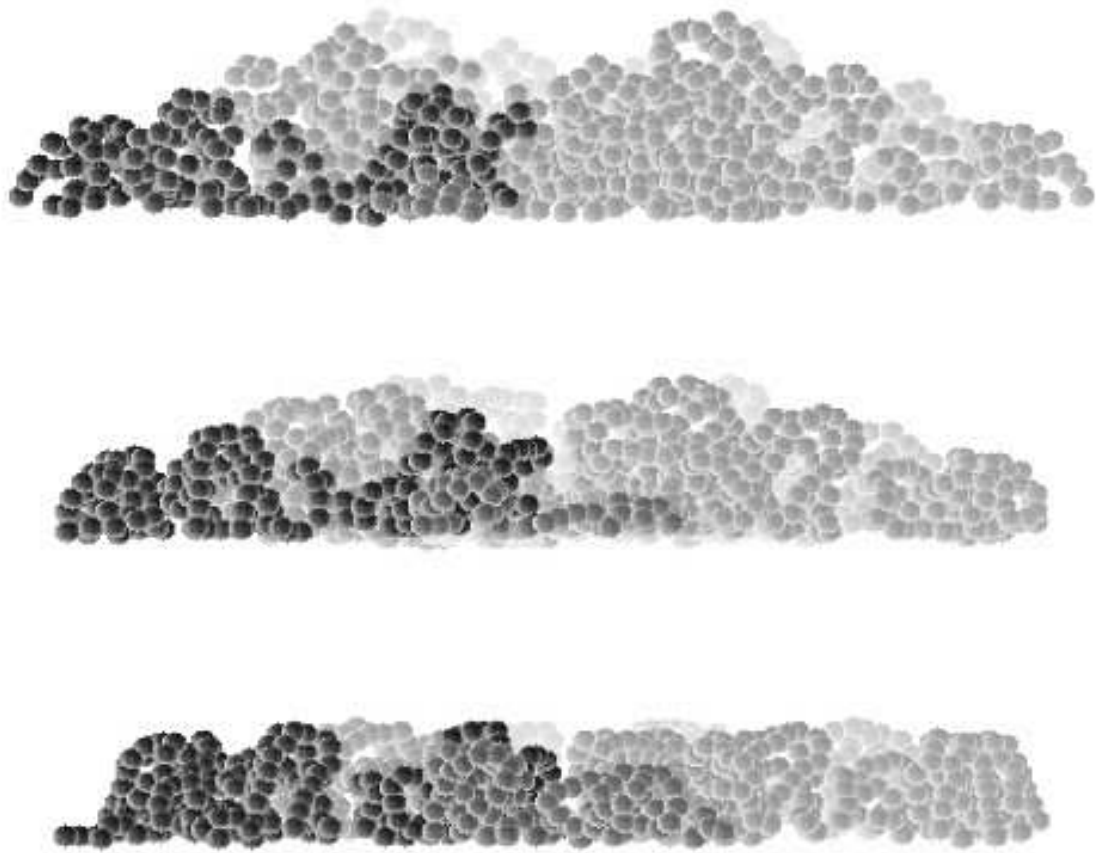}}
\caption{Snapshots of the 2000 highest C$^{\alpha}$ atoms for
$s$ equal to 284 {\AA}, 264 {\AA}, and 239 {\AA},
from top to bottom.
All of these snapshots are in the Hookean regime. }
\label{tops}
\end{figure}

\begin{figure}
\epsfxsize=6in
%\centerline{\epsffile{compres.eps}}
\centerline{\epsffile{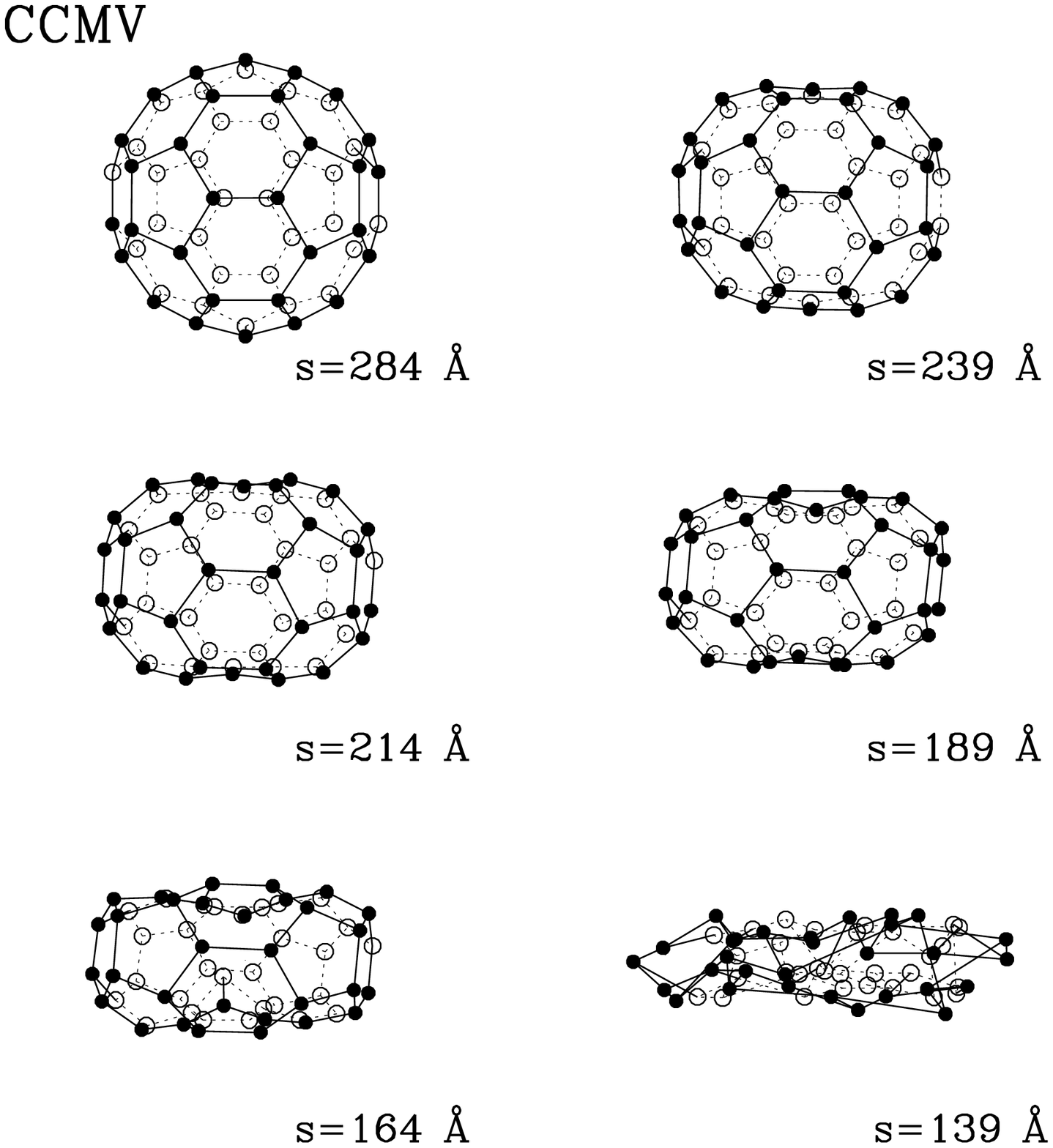}}
\vspace*{3cm}
\caption{The vertex representation of the CCMV capsid at various
stages of indentation characterized by the indicated values of $s$. The solid
circles are closer to the viewer and the open circles are in back.}
\label{vertices}
\end{figure}

In this section we focus on structural changes
in the CCMV capsid during deformation, using the trajectory
corresponding to Figure \ref{basic} to illustrate the changes.
Figures \ref{pictures}, \ref{tops}, and \ref{vertices} provide
different representations of the conformation
of the CCMV capsid at various stages of indentation.
Figure \ref{pictures} shows all C$^{\alpha}$ atoms with different gray
scales for different groups of proteins.
Snapshots show the evolution of the capsid shape from equilibrium
to one of the first states with a significant number of broken bonds,
$s=164$ \AA.
The walls first flatten and compact the region near the plates,
with the size of the flattened region growing as $s$ decreases.
The capsid remains relatively undeformed in the central region
between walls.

Figure \ref{tops} shows a closeup of the 2000 atoms that are closest
to the top plate.
In the native state, the top of the virus is relatively rough.
In the initial stages of deformation ($s=264$ \AA)
a few of the highest atoms are pushed down but only a small fraction
of the capsid surface feels the hard repulsive
potential from the walls.
The force is also very small in this range of $s$, and the stiffness
may be slightly smaller than at lower $s$ (Fig. \ref{basic}).
The number of atoms pushed up against the hard wall potential
grows as $s$ decreases.
By $s=239$ there is a pronounced flattened region and
parts of all proteins shown are in contact with the surface.

Large structural changes such as buckling and rotation are
difficult to identify in all-atom renderings.
As noted above, the capsid is constructed from triangular clusters
containing three protein chains.
A skeleton representation of the structure can be obtained by following
the center of mass of the three chains forming each triangle.
As shown in Figure \ref{vertices}, the centers of the triangles lie
at the vertices of a truncated icosahedron in the native state.
In the (110) orientation, a line connecting two vertices is at the
top and bottom of the virus.
Note that this representation makes it appear that there is a sharp
edge at the top and bottom, which would make the configuration appear
unstable against rotations.
The all-atom rendering shows that the actual surface is fairly
spherical and there is no sharp edge.

There are two pentamers and two hexamers that meet at the edges
at the top and bottom of the capsid.
As $s$ decreases from 284 {\AA} to 239 {\AA} these faces become
nearly horizontal, and by $s=214$ {\AA} many of the associated
vertices have become coplanar.
There is also a slight clockwise rotation of the capsid that is discussed below.
Further compression to $s=189$ {\AA} seems to produce buckling
with the top (bottom) edge pushed below (above) its neighbors.
These structural results are quite similar to those from finite element
calculations with a continuum model based on the mass density
of the capsid \cite{Gibbons}.
This model shows that the curvature at the top and bottom
of the capsid begins to change sign for compressions greater
than 70 {\AA}, which corresponds to $s =214$ {\AA}.
There is a pronounced region of inverted curvature at a displacement
of 100 {\AA}, which is close to the case of $s=184$ {\AA} in Figure \ref{vertices}.
It is also interesting that for this capsid orientation the continuum
calculation finds little change in the derivative of force with respect to
displacement as buckling occurs, while there is a pronounced  
change for other orientations.
Note that we did not observe softening before bond
rupture in either (100) or (111) orientations.

For $s \leq 164$ {\AA} bond breaking becomes important in
our simulations.
Fig. \ref{vertices} shows that the capsid has lost most of its structure
by $s=139$ {\AA}.
Atoms from the top and bottom of the capsid make direct contact
at these small separations, leading to a large hard core
repulsive force on the plates.
Bond breaking is not included in the continuum calculations and
one conseqeunce is that they predict a gradual decrease in
the slope of force-displacement curves as $s$ decreases to $164$ {\AA},
while our calculations show a precipitous drop in force at this point.

\begin{figure}
\epsfxsize=6in
%\centerline{\epsffile{bbroken.eps}}
\centerline{\epsffile{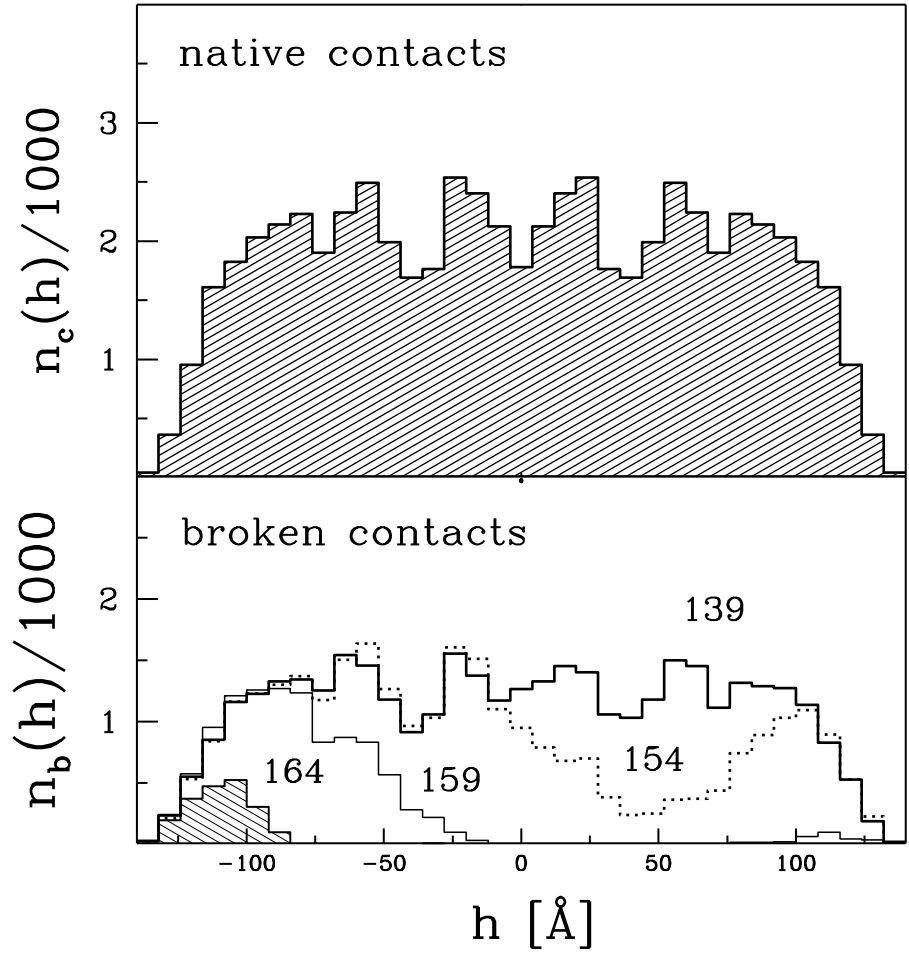}}
\vspace*{3cm}
\caption{
The top panel shows the $z$-dependent distribution of the native contacts, $n_c(h)$,
as explained in the text. The bottom panel shows the distributions of contacts
that are broken at the indicated values of $s$ in {\AA}.
}
\label{ruptured}
\end{figure}

Figure \ref{ruptured} illustrates the spatial distribution of broken bonds
for the simulation shown in Figure \ref{vertices}.
Each native contact is characterized by its initial
"height" $h$, or more precisely, the $z$-coordinate of its  
geometrical center in the native conformation.
The top panel shows the height distribution of native contacts.
The bottom panel shows the number with each initial height that are broken
at various plate separations.
Bond breaking initiates just before $s$ decreases to 164 {\AA},
which is just at the end of the linear portion of the force-displacement
curve in Figure \ref{basic}.
Breaking initiates near the bottom of the virus
for this trajectory,
but failure initiates at the top in other trajectories.
As seen in Fig. \ref{vertices} and continuum calculations,
stresses and strains are largest in these extremal regions.

At $s=164$ {\AA} about 2000 contacts are broken.
Decreasing $s$ by just 5 {\AA} ruptures nearly all of the bonds
that eventually fail in the bottom region and initiates
failure at the top of the capsid.
A further decrease by 5 {\AA} has spread the failure over most of the
capsid.
The number of broken bonds then saturates at lower $s$ as
shown in Figure \ref{contacts}.
The saturated distribution is illustrated by the results for $s=139$ {\AA} in
Figure \ref{ruptured}.
Note that the distribution mirrors the number of native contacts,
indicating that a fairly constant fraction of the initial bonds has broken
throughout all regions of the capsid.
As noted above, almost all the bonds binding different proteins together
have failed, leading to complete loss of the capsid structure.

\begin{figure}
\epsfxsize=6in
%\centerline{\epsffile{bnwelda.eps}}
\centerline{\epsffile{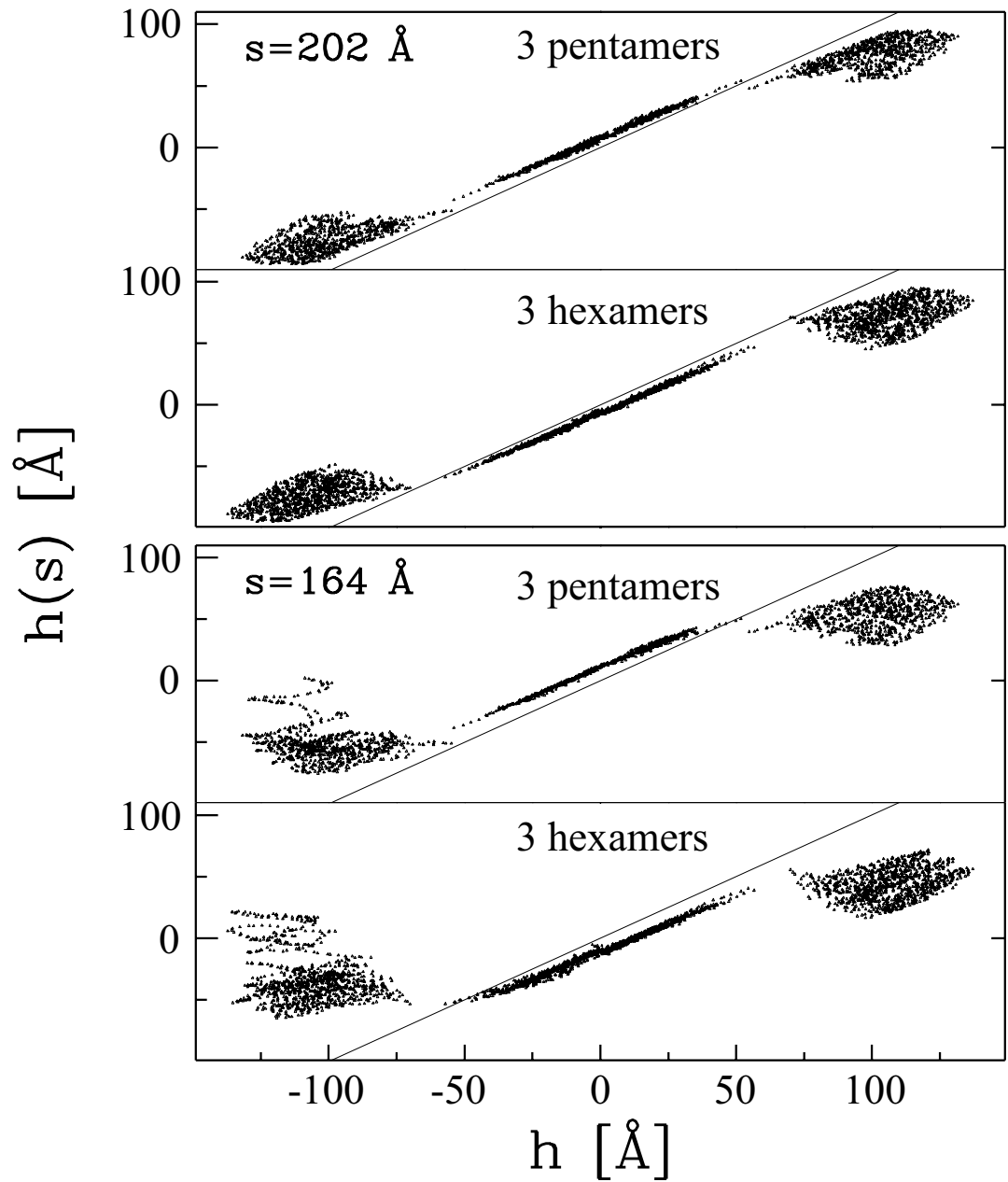}}
\vspace*{3cm}
\caption{Scatter plots of final heights of C$^{\alpha}$ atoms
as a function of initial heights in the native structure.
The top two panels are for $s$=202 {\AA} and the bottom two panels are
for $s$=164 {\AA}.
Hexamers and pentamers were selected to represent the polar and equatorial
regions.}
\label{pentax}
\end{figure}

We now focus on quantifying the distribution of deformation within the capsid,
by considering
the $z$-displacements of individual atoms.
They can be illustrated directly 
through a scatter plot in which the
atoms' $z$-coordinates at separation $s$, $h(s)$, are plotted against
their native $z$-position, $h$.
Figure \ref{pentax} shows such scatter plots for the
C$^{\alpha}$ atoms belonging to selected 
pentamers (3 out of 12) and hexamers (3 out of 20).
In each case the n-mers are chosen to represent the top, center
and bottom and to prevent overlap of points in the scatter plot.
Results are shown for $s=202$ {\AA} and $s=164$ {\AA}
where bond breaking becomes evident.

At both plate separations the central n-mers are nearly undeformed.
The final height differs from the initial height (solid line)
in two ways.
One is a shift up or down that reflects the small clockwise rotation
mentioned in regard to Figure \ref{vertices}.
The second is a small change in slope.
For both plate separations the slope is 0.93 for the pentamer,
while for the hexamer the slope decreases from 0.96 to 0.93 with decreasing $s$.
The decrease in slope from unity results mainly from compression, but is also
affected by rotation of the n-mers.
The mean rotation of the capsid is only about 4$^\circ$ for $s=202$ {\AA}.

The behavior of the outer n-mers is very different.
In each case, the largest values of $|h|$ are very close to the height of
the walls, $s/2$.
As these outer atoms are pushed inwards, they displace the entire thickness
of the capsid with them.
This leads to a broadening of the scatter plots by an amount
comparable to the capsid thickness.
Note that the envelope corresponding to the largest values of $|h|$
and the outer edge of the capsid does not merge smoothly with the 
lines for the central n-mers.
This reflects a rapid change in the surface normal at the edge of the
central n-mers that is also evident in Figure \ref{vertices}.

To see if the deformations of the capsid in our simulations are consistent
with continuum models, we compared them to the finite-element results of
Gibbons and Klug \cite{Gibbons}.
Since the finite element nodes do not correspond to individual atoms,
horizontal slices of width 1 {\AA} at selected heights were chosen 
and nodes at the inner and outer edges of the capsid were followed.
For the finite-element model these are the inner and outer
nodes of the mesh and for the molecular model they are the inner
and outer C$^\alpha$ atoms.
The continuum calculations moved just one wall so that the center
of mass moved.
We have corrected for this so that zero height corresponds to the
center of mass in both cases.

\begin{figure}
\epsfxsize=6in
%\centerline{\epsffile{n7hequat.eps}}
\centerline{\epsffile{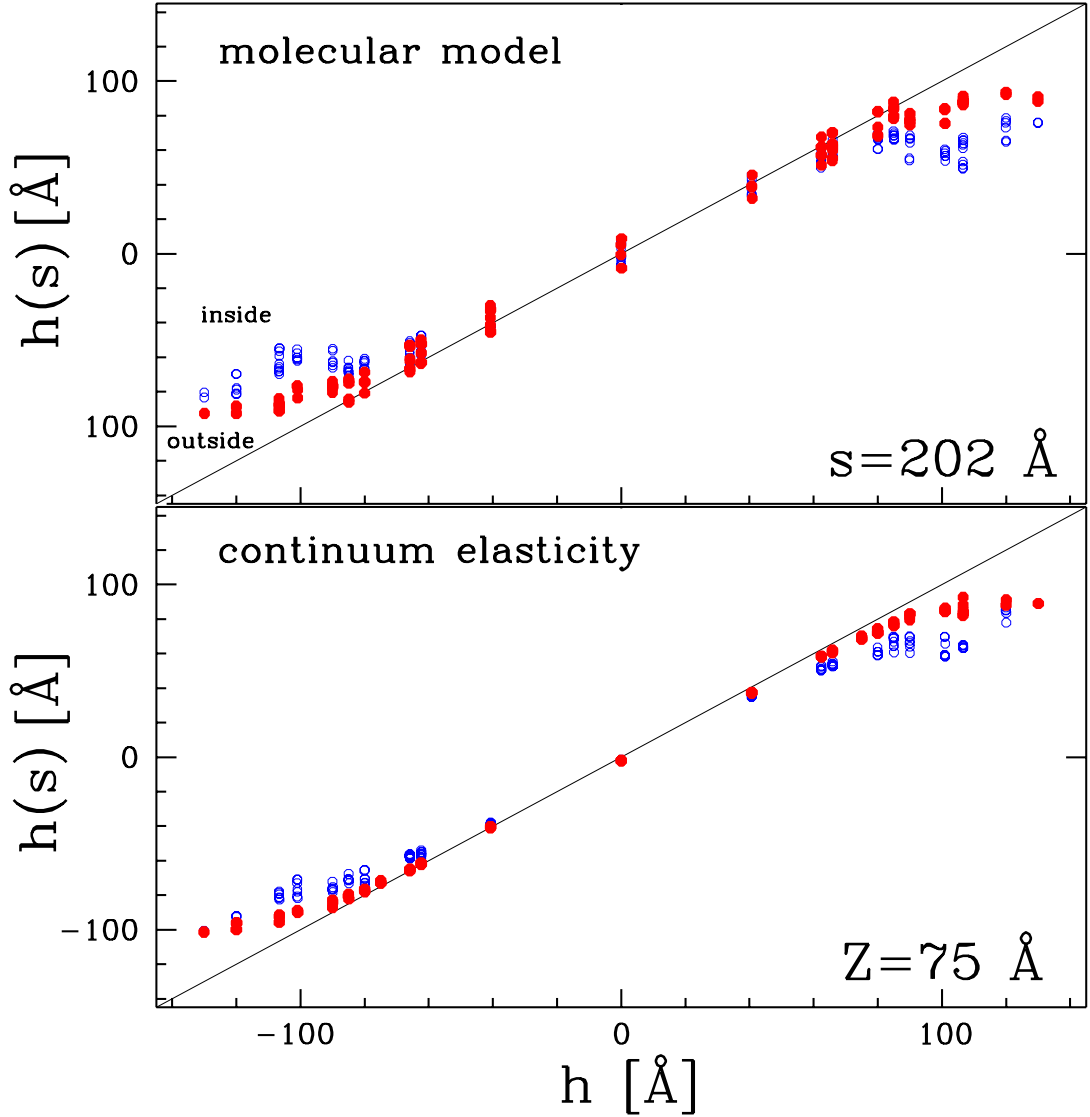}}
\vspace*{3cm}
\caption{Scatter plots of final vertical positions as a function of
initial height for outer (solid symbols) and inner (open symbols)
boundaries of the capsid.
The top panel shows postions of C$^{\alpha}$ atoms for $s$=202 {\AA},
and the bottom panel shows node positions from the continuum model of
Gibbons and Klug at a wall displacement of $Z=75$ {\AA}.
\cite{Gibbons}.}
\label{cont202}
\end{figure}

\begin{figure}
\epsfxsize=6in
%\centerline{\epsffile{nhequat.eps}}
\centerline{\epsffile{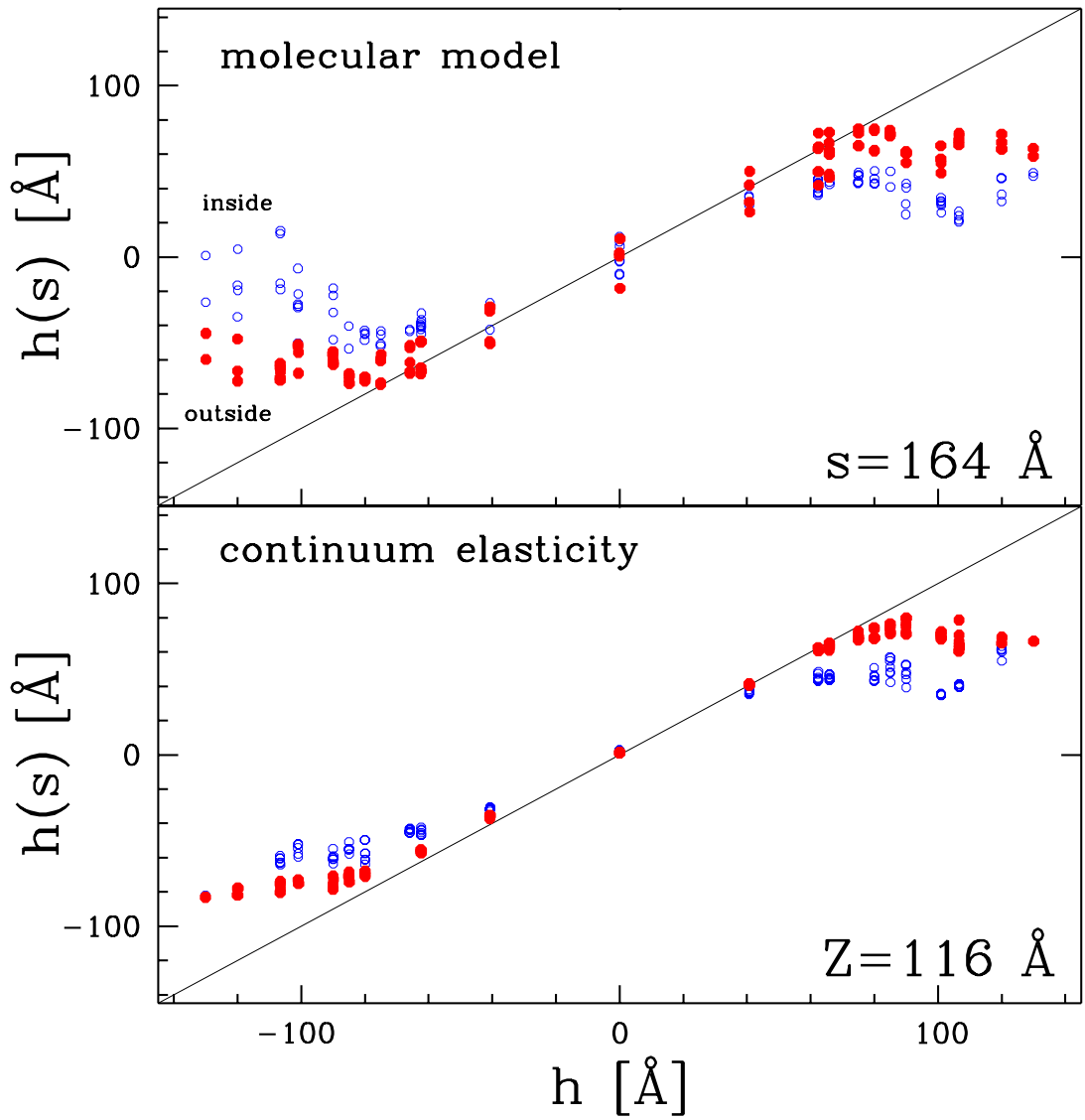}}
\vspace*{3cm}
\caption{Similar to Figure \ref{cont202} but for
a higher compression corresponding to $s$=164 {\AA}.}
\label{cont164}
\end{figure}

Figures \ref{cont202} and \ref{cont164} compare the finite-element and
molecular models for two plate separations.
Figure \ref{cont202} shows molecular model results for $s=202$ {\AA} which
corresponds to a compression of $Z=82$ {\AA}.
These results are compared to finite element results at $Z=75$ {\AA}.
Figure \ref{cont164} presents results for $Z=116$ {\AA} in the
continuum model and $Z=120$ {\AA} in the molecular
model, which corresponds to the onset of bond breaking ($s=164$ {\AA}).

For both separations the molecular results show a spread in final
heights for atoms with the same small initial height
\footnote{It should be noted
that the widths of the scatter plots in the equatorial regions shown in
Figure \ref{pentax} are smaller than those shown in the top panels of
Figures \ref{cont202} and \ref{cont164}. This is  simply because only single
equatorial hexamers and pentamers are displayed in the previous figure
and all n-mers are considered now.
}.
This reflects the symmetry breaking rotation of the capsid that is
evident in Figures \ref{vertices} and \ref{pentax}.
Symmetry breaking is suppressed in the continuum model by the lack of
thermal fluctuations and the fact that a friction force was imposed to
suppress sliding of the virus relative to the wall.
As shown in Fig. \ref{pentax},
the spread in results for the molecular model results from
coherent shifts in the height of each n-mer through a relatively
rigid rotation.
For each n-mer there is a linear relation between initial and final height.
The continuum model shows a linear relation for all nodes with
slightly different slopes for the inner and outer atoms.
A least mean squares fit to heights of all nodes between $\pm 40$ {\AA}
gives 0.90 for the continuum model at $Z=75$ {\AA}.
The same fit for atoms in the molecular model gives a slope of 0.94,
indicating that there is significantly less strain in the central region.

The deviations between continuum and molecular models are slightly larger
near the top and bottom of the capsid.
Both show a spread between inner and outer edges, because the outer
surface is in contact with the wall and the inner surface remains
at a fixed distance below the outer surface.
The continuum model gives a smoother curve for the outer surface,
while the outer height appears to vary non-monotonically in the
molecular model.
A local dip in final height at an initial height of around 100 {\AA} 
is visible for both wall separations.
The region where the final height corresponds to contact with the
wall also extends over a larger range of heights in the molecular model.
This is consistent with less strain being accommodated in the central
region, and more in the regions near the walls.

\begin{figure}
\epsfxsize=6in
%\centerline{\epsffile{strain.eps}}
\centerline{\epsffile{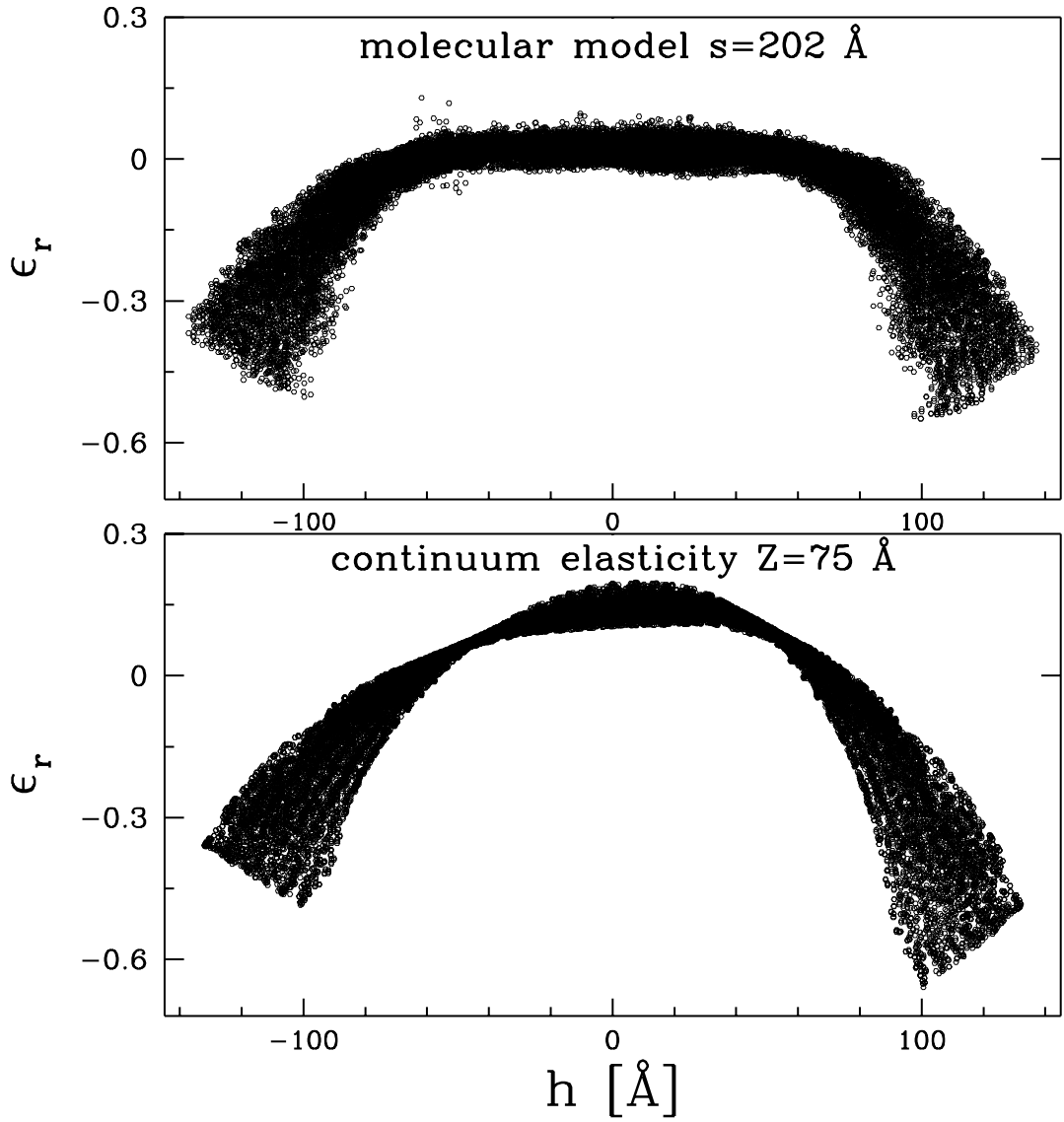}}
\vspace*{3cm}
\caption{Scatter plots for the radial strain as a function of height
corresponding to a wall separation of $s=202$ {\AA}.
The top panel is for the molecular model and the bottom from
the data provided by Gibbons and Klug
from a continuum model \cite{Gibbons}.
}
\label{newplot}
\end{figure}

The radial strain for both models was also calculated.
Defining $\vec{r}$ as the initial position relative to the
center of mass and $\vec{dr}$ as the change in this vector,
the radial strain $\epsilon_r = \vec{r}\cdot \vec{dr}/\vec{r}\cdot\vec{r}$.
Figure \ref{newplot} contrasts the molecular and continuum response for
$s=202$ {\AA} and $Z=75$ {\AA}, respectively.
In both cases, the strain is negative at large $|z|$ where compression
has flattened the top and bottom.
In the central region the capsid bows out and $\epsilon_r$ is positive.
For the continuum case the inner and outer regions form smooth
parabolas that cross at intermediate $z$.
There is some symmetry breaking due to the onset of buckling on one side.
In the central region, the radius expands by an average of about 15\%.
The molecular model shows substantially less expansion, with
an average of less than 5\%.
The envelope of the data is also less smooth, showing a nearly
constant expansion in the center and a rapid change in slope for
$|z| > 60$ {\AA}.
This is consistent with the apparent change in slope in
Figure \ref{pentax}
and may reflect localized bending at
joints between proteins in this region.
Note that Fig. \ref{cont164} shows that the flattened
region of the capsid has extended down to atoms with
initial heights of order 60 {\AA} by $s=164$ {\AA}.
It may be that this leads to failure
because strain can no longer be accommodated
by the outer regions.

\begin{figure}
\epsfxsize=6in
%\centerline{\epsffile{pendisp.eps}}
\centerline{\epsffile{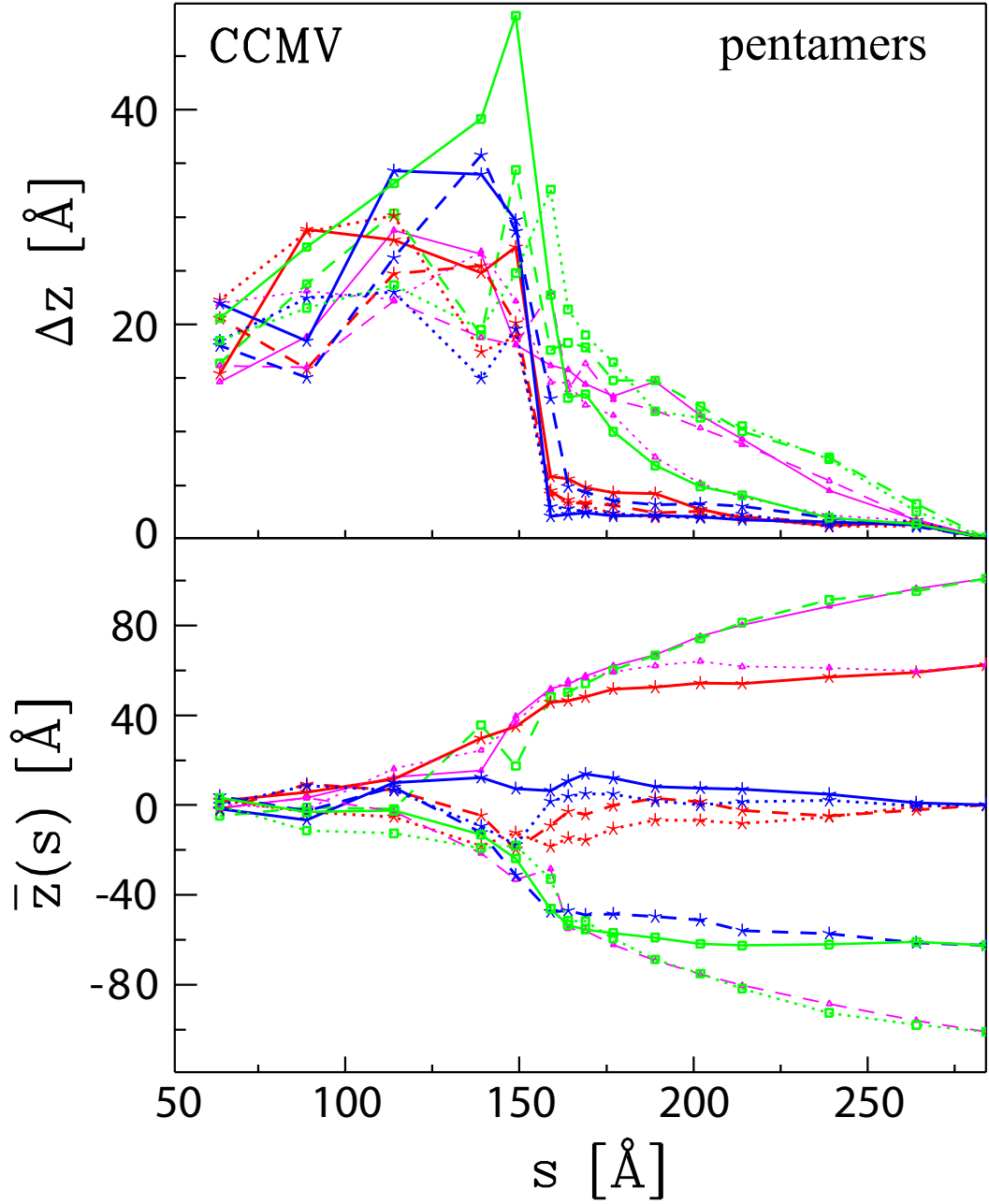}}
\vspace*{3cm}
\caption{The bottom panel shows the height of the centers
of mass of atoms belonging to 12 individual pentamers.
The top panel shows rms variation in the height change
of atoms belonging to the same pentamer.
The outer lines in the bottom panel correspond to the polar pentamers
and the lines in the middle to the equatorial regions.
The same symbols are used in the top panel.
}
\label{pentrms}
\end{figure}

\begin{figure}
\epsfxsize=6in
%\centerline{\epsffile{nhexdisp.eps}}
\centerline{\epsffile{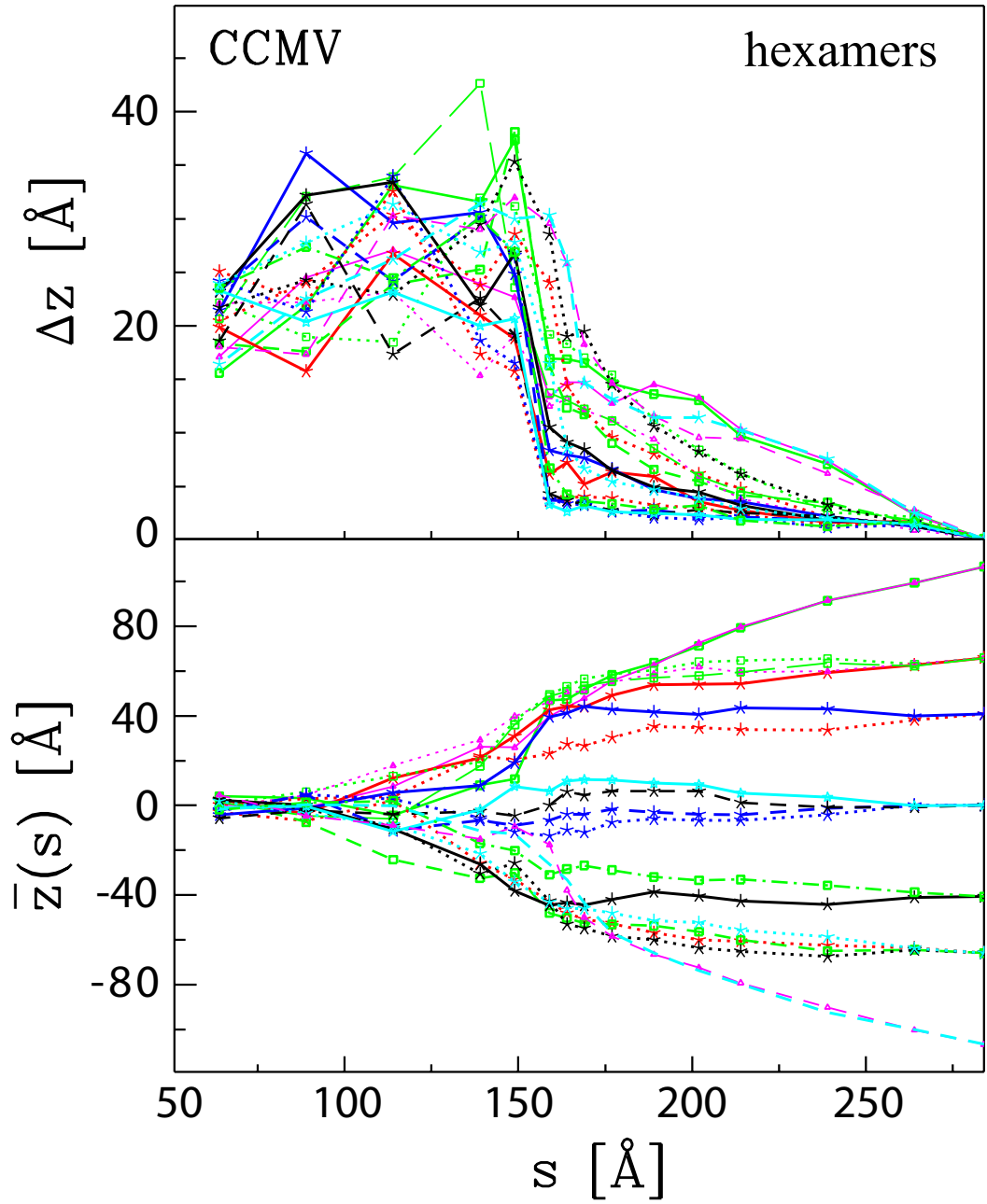}}
\vspace*{3cm}
\caption{Similar to Figure \ref{pentrms} but for the 20 hexamers.
}
\label{hexrms}
\end{figure}

To illustrate the development of deformations with decreasing separation,
the center of mass height $\bar{z}$ and rms variation $\Delta z$ was
calculated for atoms comprising each of the n-mers.
Figures \ref{pentrms} and \ref{hexrms} show results for all pentamers and
hexamers, respectively.
The locations of the centers of mass indicate a fairly smooth
transition to a flattened "sandwich" state as the wall separation
becomes comparable to the thickness of capsid walls.
At large $s$, the height of the top and bottom n-mers changes much more
rapidly than those in the center,
indicating yet again that strain is concentrated in the regions near
the walls.
The mean displacement of n-mers at the same intermediate height shows
a relatively smooth splitting, that is associated with the symmetry
breaking rotation seen before.
The splitting is not sudden and is reversible when the load is removed.
The mean rotation is about 4$^\circ$ for $s=202$ {\AA}.

Note that one of the n-mers in each pair stays at a nearly constant
height while the other n-mer moves linearly towards the center.
In contrast, the n-mers that start in the center move away from the
center with symmetry related pairs displacing by equal and opposite
amounts.
Once significant bond-breaking sets in, the heights change rapidly.
Note that the relative height of outer and more central n-mers changes
sign, indicating buckling.
It is also interesting that the centers of the outer n-mers become
equal to the centers of more central n-mers at the point where
bond breaking sets in.
These heights are also equal to the height where rapid bending
is evident in Fig. \ref{newplot}.

The trends in the rms height change mirror the shifts in the mean.
Initially there is a fairly linear increase in $\Delta z$ with the
greatest slope for the n-mers farthest from the equator.
The rms displacements are very large, comparable to the mean motion
in many cases.
This reflects the rotation of the outer n-mers as they move into contact
with the wall.
When bonds begin to break, all of the n-mers show rms variations
of 20 to 40 {\AA} which is comparable to the thickness of the capsids
and to the plate separation at the smallest $s$.
The n-mers are made up of several proteins that are no longer bonded
together at small $s$ and thus deform incoherently.

\section{Summary and Conclusions}

A coarse-grained molecular model was used to study the mechanical response
of CCMV and CPMV.
Both show a linear elastic response at small deformations followed
by a rapid drop in force.
The response is reversible in the elastic regime and becomes irreversible
after the force drop due to rupture of bonds within the capsid.
Although the two capsids have similar radii, widths and densities,
their mechanical response is dramatically different.
CPMV is an order of magnitude stiffer and fails at a much smaller
compression and much larger peak force.

The dramatic differences between CCMV and CPMV are the result of a relatively
modest change in the number of bonds.
Insight to this dependence comes from comparing the response to elastic shell
models.
While the effective modulus from fits to CPMV is an order of magnitude larger
than that for CCMV, it is more than an order of magnitude smaller
than that of an fcc crystal with the same bonding energies.
The crystal would also fail at much smaller deformations.
The mean number of neighbors drops from 12 in the fcc crystal
to 8 in CPMV and only 6 in CCMV.  Four are required to provide mechanical
stability in 3 dimensions and studies of rigidty percolation
show that the modulus is a strong function of the number and
distribution of neighbor bonds.
Thus viruses may be able to alter their stiffness and ability
to accommodate large strains through small variations in the
number of interprotein bonds.
It would be interesting to perform experiments on CPMV and other
virus capsids to see if the effective modulus does vary by
an order of magnitude and whether this mechanical property
is correlated to the
environment or biological functions of the different capsids.

Elastic continuum models produce a nonlinear 
drop in force when the shell buckles.
Since there are no thermal fluctuations, buckling occurs
at a well-defined displacement where the system becomes
linearly unstable.
In the molecular model the force drop is associated
with rupture of almost all the interprotein bonds, allowing
the capsid to fragment.
The capsid did not reform on the time scale of our simulations, but
experiments show some recovery over times of order minutes \cite{Michel}.

The simulation results are consistent with the force drop
being an activated transition where
breaking of a few bonds leads to stress transfer and a cascade of
additional bond breaking.
In particular, different trajectories at the same deformation rate
follow a common force curve before and after the transition, but
the transition occurs at different displacements.
As the wall separation is decreased more slowly, there
is more time for thermal activation and the transition
occurs at lower forces and displacements.
Experiments show a similar trend in the onset of irreversibility with
velocity \cite{Michel}, while there is no shift in the location
of the buckling transition in deterministic continuum models.
As is typical of activated transitions,
the stiffness in the elastic region is relatively insensitive
to deformation rate \cite{eyring}.
An unusual activated rate dependence of the stiffness was observed in
recent all-atom simulations with a much smaller
probe and higher velocities \cite{zink09}.
As pointed out by the authors of this study, the rate
dependence they observe suggests that their deformation rates
were too high to allow the virus structure to relax completely.

Quantitative comparison to experiments is complicated by rate dependence
and uncertainties in the binding energy within our molecular model.
The latter also leads to uncertainties in the appropriate temperature.
Our slowest rates are two to four orders of magnitude faster than
experimental rates and the wall
displacement that produces irreversibility ranges from 150 to
90 {\AA}.
The range of rates and statistics are too small to extrapolate these results
to the displacements of about 40 {\AA} where irreversibility
is observed in experiments.
Continuum models give buckling at roughly twice this displacement,
but buckling does not produce a strong drop in the force, particularly
for the (100) orientation.
The molecular model gives buckling at about the same displacement
as the continuum model and also shows that it does not have
a strong impact on the stiffness.
The sharp force drop in the molecular
model occurs at larger deformations when bonds begin to break.
At the rates considered here, bonds break after the buckling transition,
and the onset of breaking shifts relatively rapidly with velocity.
One might expect that buckling produces highly strained bonds that are
more susceptible to activated rupture and that the velocity dependence
might become much slower when the transition occurs at the smaller
displacements found in experiments.

Experiments on CCMV at pH=5 give a peak force of $F_m=0.60\pm 0.04$nN
and a stiffness (-dF/ds) of $k =0.15\pm 0.01$ N/m in the elastic region
\cite{Michel}.
In our simulations, $k \approx 0.05$ $\epsilon$/{\AA}$^2$ and the peak force
is about 4 $\epsilon$/{\AA} at the lowest velocity.
Using a binding energy of $\epsilon=110$ pN{\AA},
these values correspond to $k \sim .055$ N/m
and $F_m \sim 0.4$ nN, which are of the same order of magnitude as 
measured values.
It is not surprising that agreement is better for $F_m$ since $\epsilon$
was obtained from fits of simulated force curves during mechanical
unfolding to experiments that, as here, were done at lower rates.
Increasing $\epsilon$ would increase both $F_m$ and $k$, but lowering
the rate would bring $F_m$ back toward the experimental value
and also decrease the displacement at failure toward the 40 {\AA}
found in experiment.
A larger $\epsilon$ would also decrease the effective temperature,
which also increases $k$ and $F_m$.
Thus the quantitative agreement is as good as can reasonably expected
from a coarse-grained model.

The local deformations within the CCMV capsid were compared to continuum
calculations.
While the continuum model captures the gross features of the deformation
there are deviations that must reflect variations in the local elastic
properties due to the distribution of native contacts.
In the molecular model,
more of the deformation is accommodated in the top and bottom of the capsid.
The central region is compressed less, and also shows substantially less
radial expansion than the continuum model.
There is a relatively large change in angle at a height of about 60 {\AA}
that suggests bending.
Given the other evidence that the regions between proteins are less
rigid, it seems likely that bending is localized between proteins.
More detailed studies of local deformations and stresses is in progress.

\acknowledgments
Discussions with R. Bruinsma, W. Klug, and A. Zlotnick are warmly
appreciated. We thank W. S. Klug and M. M. Gibbons for providing us with
results from their continuum model.
This work has been supported by the grant N N202 0852 33 from the Ministry
of Science and Higher Education in Poland, the EC FUNMOL project under
FP7-NMP-2007-SMALL-1, and by the European Union within European
Regional Development Fund, through grant Innovative Economy
(POIG.01.01.02-00-008/08).
Partial support for this work was also provided by the National Science
Foundation under Grant No. DMR-0454947.

%%%%\newpage

\vspace*{2cm}
\centerline{FIGURE CAPTIONS}

%FIGURE 1

%FIGURE 2

%FIGURE 3

%FIGURE 4

%FIGURE 5

%FIGURE 6

%FIGURE 7
\begin{figure}
\epsfxsize=6in
%\centerline{\epsffile{sbszar.eps}}
\centerline{\epsffile{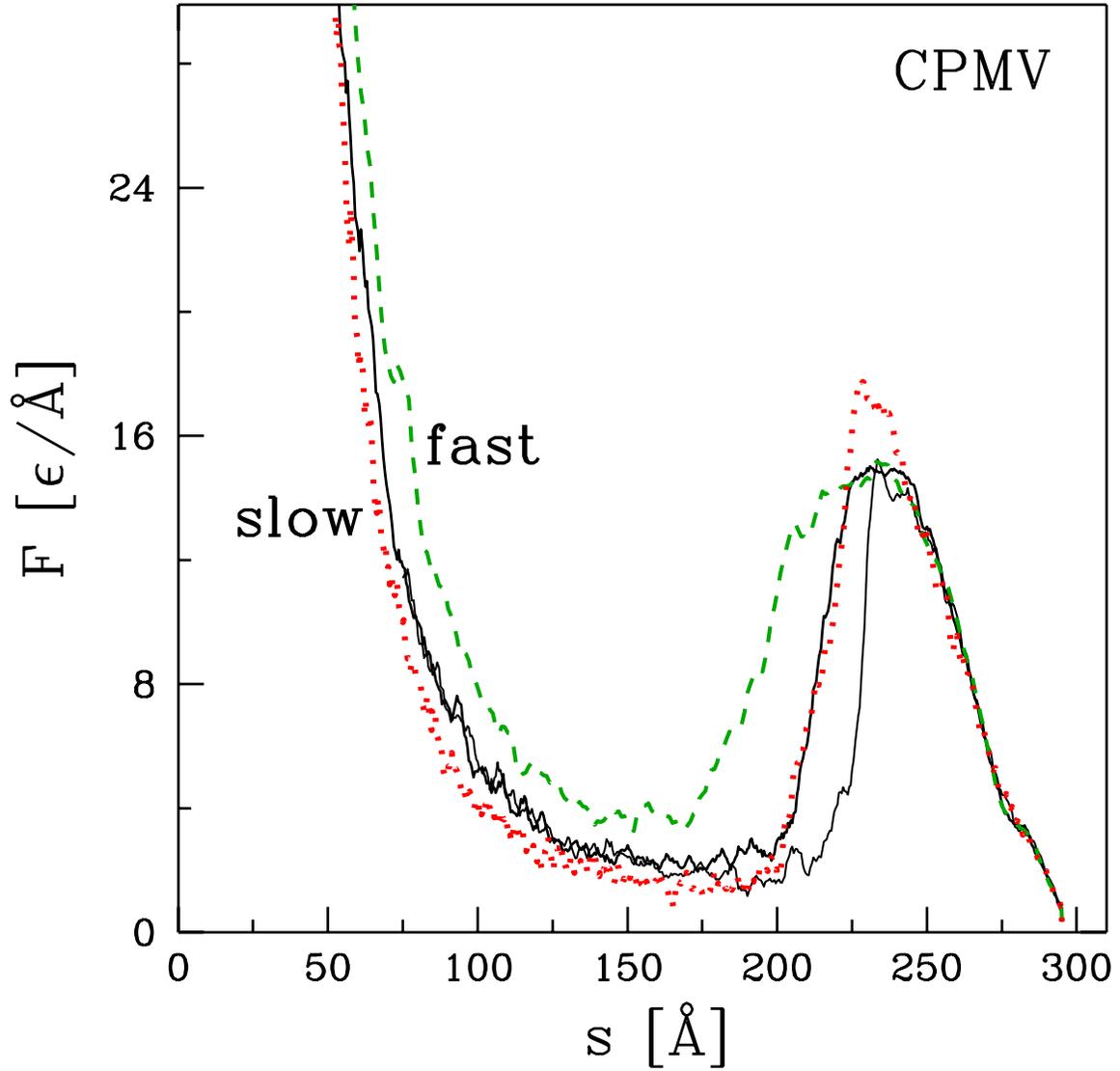}}
\vspace*{3cm}
\caption{Similar to Figure \ref{speed} but for the model CPMV. Two trajectories
are shown for $v_p=0.0025 $ {\AA}/$\tau$ (solid lines) and single trajectories
at half (dotted line) and twice (dashed line) this speed.
}
\label{speedcpmv}
\end{figure}

%FIGURE 8

%FIGURE 9

%FIGURE 10

%FIGURE 11

%FIGURE 12

%FIGURE 13

%FIGURE 14

%FIGURE 15

%FIGURE 16

%FIGURE 17

%FIGURE 18

\end{document}